\documentclass[prb, twocolumn,floatfix,superscriptaddress]{revtex4}

\usepackage{graphicx}

\usepackage{verbatim}

\usepackage{times}

\usepackage{subfigure}

\usepackage{amsmath}

\usepackage{natbib}


\begin{document}

\title{Pressure-energy correlations in liquids. II. Analysis and 
consequences}

\author{Nicholas P. Bailey}
\email{nbailey@ruc.dk}
\affiliation{DNRF Center ``Glass and Time'', IMFUFA, Dept. of Sciences, 
Roskilde University, P.O. Box 260, DK-4000 Roskilde, Denmark}

\author{Ulf R. Pedersen}
\affiliation{DNRF Center ``Glass and Time'', IMFUFA, Dept. of Sciences, 
Roskilde University, P.O. Box 260, DK-4000 Roskilde, Denmark}

\author{Nicoletta Gnan}
\affiliation{DNRF Center ``Glass and Time'', IMFUFA, Dept. of Sciences, 
Roskilde University, P.O. Box 260, DK-4000 Roskilde, Denmark}

\author{Thomas B. Schr{\o}der}
\affiliation{DNRF Center ``Glass and Time'', IMFUFA, Dept. of Sciences, 
Roskilde University, P.O. Box 260, DK-4000 Roskilde, Denmark}

\author{ Jeppe C. Dyre }
\affiliation{DNRF Center ``Glass and Time'', IMFUFA, Dept. of Sciences, 
Roskilde University, P.O. Box 260, DK-4000 Roskilde, Denmark}

\begin{abstract}
We present a detailed analysis and discuss consequences 
of the strong correlations of the configurational parts of pressure and 
energy in their equilibrium fluctuations at fixed volume
reported for simulations of several liquids in the previous (companion)
paper [arXiv:0807.0550].
The analysis concentrates specifically on the single-component
Lennard-Jones system. We demonstrate that the potential may be replaced, at 
fixed volume, by an effective power-law, but not simply because only short
distance encounters dominate the fluctuations. Indeed, contributions to the
fluctuations are associated with the whole first peak of the radial 
distribution function, as we demonstrate by an eigenvector analysis of the
spatially resolved covariance matrix. The reason the effective power-law works 
so well depends crucially on going beyond single-pair effects and on the 
constraint of fixed volume. In particular, a better approximation to the 
potential includes a linear term, which contributes to the mean values of
potential energy and virial, but not to their fluctuations, 
for density fluctuations which conserve volume.
We also study in considerable detail the zero temperature limit of the
(classical) crystalline phase, where the correlation coefficient
becomes very close,
but not equal, to unity, in more than one dimension; in one dimension the 
limiting value is exactly unity. 
In the second half of the paper we consider four consequences of 
strong pressure-energy correlations: 
(1) analyzing experimental data for supercritical Argon we find 
96\% correlation;
(2) we discuss the particular significance acquired by the
correlations for viscous van der Waals liquids 
approaching the glass transition: For strongly correlating viscous
liquids knowledge of just one of the eight frequency-dependent 
thermoviscoelastic response functions basically implies knowledge of them all;
(3) we 
re-interpret aging simulations of ortho-terphenyl carried out by
Mossa {\it et al.} in 2002, showing their conclusions follow
from the strongly correlating property; and (4) we briefly discuss the
 presence of the correlations (after appropriate time-averaging) in model
biomembranes, showing that significant correlations
 may be present even in quite complex systems.
\end{abstract}

\newcommand{\angleb}[1]{\langle #1 \rangle}
\newcommand{\nod}{\noindent}
\newcommand{\half}{\frac{1}{2}}
\newcommand{\bfa}[1]{\mathbf{#1}} 

\date{\today}

\maketitle

\section{Introduction}

In the companion paper\cite{Bailey/others:2008b} to this work, referred to as 
Paper I, we detailed
the existence of a strong correlation between the configurational parts of
pressure and energy in several model liquids. Recall that (instantaneous) 
pressure $p$ and
energy $E$ have contributions both from particle momenta and
positions:\cite{Allen/Tildesley:1987}

\begin{align}
  p &= Nk_BT(\bfa{p}_1,\ldots,\bfa{p}_N)/V+W(\bfa{r}_1,\ldots,\bfa{r}_N)/V \\
  E &= K(\bfa{p}_1,\ldots,\bfa{p}_N) + U(\bfa{r}_1,\ldots,\bfa{r}_N),
\end{align}

\nod where $K$ and $U$ are the kinetic and potential energies, respectively,
and $T(\bfa{p}_1,\ldots,\bfa{p}_N)$ is the
``kinetic temperature'',\cite{Allen/Tildesley:1987}
proportional to the kinetic energy per particle. The configurational 
contribution to pressure is the virial $W$, which for a translationally
 invariant potential energy function $U$ is defined\cite{Allen/Tildesley:1987}
by

\begin{equation}\label{generalWformula}
W = -\frac{1}{3} \sum_i \bfa{r}_i \cdot \bfa{\nabla}_{\bfa{r}_i} U
\end{equation}

\nod  where $\bfa{r}_i$ is the position of the $i$th particle. Note that $W$ 
has dimension energy. In the case of a pair potential 
$U_{\textrm{pair}} = \sum_{i<j} v(r_{ij})$ the expression for the virial
becomes\cite{Allen/Tildesley:1987}

\begin{equation}
W_{\textrm{pair}} = -\frac{1}{3}\sum_{i<j} r_{ij} v'(r_{ij}) = -\frac{1}{3} 
\sum_{i<j} w(r_{ij})
\end{equation}

\nod where $w(r)\equiv rv'(r)$. 

It is the correlation between $U$ and $W$ that 
we are interested in, quantified by the correlation coefficient

\begin{equation}
  R=\frac {\langle\Delta W \Delta U\rangle}{\sqrt{\langle(\Delta
      W)^2\rangle}\sqrt{\langle(\Delta U)^2\rangle}}.
\end{equation}

\nod Paper I documented the correlation in many systems, 
showing that this is often quite strong, with correlation coefficient $R>0.9$, 
while in some other cases
it was found to be weak or almost non-existent. The latter included
in particular models with additional 
significant Coulombic interactions. The purpose of 
this paper is two-fold. First we give a comprehensive
analysis of the source of the correlation in the simplest ``strongly
correlating'' model liquid, the standard single-component Lennard-Jones (SCLJ) 
fluid. Paper I presented briefly an
explanation in terms of an effective inverse power-law potential. Here we 
elaborate on that in greater detail, and go beyond it. Secondly we discuss a
few
observable consequences and applications of the strong correlations. These
range from their measurement in a real system to applications to 
systems as diverse as supercooled liquids and biomembranes.

In section~\ref{Analysis} we present a detailed analysis
for the SCLJ case, first in terms of an effective inverse 
power-law with exponent
$\sim 18$. This accounts for the correlation at the level of individual
pair-encounters by assuming that only the repulsive part of the potential,
corresponding to distances less than the minimum of the potential $r_m$, is
relevant for fluctuations, and that this may be well approximated by an inverse
 power-law. The value 18 is significant since this explains the 
``slope'' $\gamma$ defined as

\begin{equation} \label{gamma_defined}
\gamma \equiv \sqrt{\frac{\angleb{(\Delta W)^2}}{\angleb{(\Delta U)^2}}},
\end{equation}

\nod observed to be $\sim 6$ for Lennard-Jones systems (Paper I). The slope is 
exactly $n/3$ for a pure inverse power-law potential with exponent $n$, a case
with perfect $W,U$ correlation (Paper I).
Section~\ref{Analysis} continues with a discussion of the SCLJ crystal, which 
is also strongly correlating. This would seem to invalidate the dominance of
the repulsive part, since only presumably distances around and beyond the 
potential minimum are important, at least at low and moderate pressure.
In this case the correlation
can be explained only when summation over all pairs is taken into account, thus
the correlation emerges as a collective effect.
There is a connection between the slope obtained in this way and that given 
by the effective inverse power-law; in fact they are quite similar. 
The third subsection in section~\ref{Analysis}
gives a more systematic analysis of which regions dominate the fluctuations in
the liquid phase,
using an eigenvector decomposition of the spatially-resolved covariance
matrix. This matrix represents the contributions to the (co-)variances of 
potential energy and virial from different pair separations. It is demonstrated
 that the region around the minimum of the potential plays a substantial role.
The final subsection in section~\ref{Analysis} provides a synthesis of the 
insights from the previous subsections, resulting in an ``extended effective
power-law approximation'', which includes a linear term. The main point is that
a linear term in the pair potential will contribute to the mean values, but
not to fluctuations, of $W$ and $U$, if the volume is constant.

In section~\ref{Consequences} we discuss some consequences, starting
by considering whether the instantaneous correlations can be related to a 
measurable quantity in the normal liquid state, and demonstrating this with
data for supercritical Argon, finding a correlation coefficient of 0.96.
Next we focus on
consequences for highly viscous liquids, where time-scale separation implies
that instantaneous correlation between virial and potential
energy can be related to a correlation between the time-averaged pressure and
energy. The third subsection discusses consequences for aging, while the 
fourth briefly
discusses connections with recent work by Heimburg and Jackson on biomembranes.

Finally, section~\ref{Conclusion} concludes with an outlook
reflecting on the broader significance of strong correlation, 
and its implications for the understanding of liquids,  
particularly in the context of viscous liquids (which has
been our main motivation throughout this work).

\section{\label{Analysis}Analysis}

\subsection{\label{effectivePowerLaw}The effective inverse power law}

In this section we consider the SCLJ system, 
in the hope that it is simple enough that a fairly complete understanding of 
the cause
of strong correlations is possible. Recall that $R>0.9$ for a wide range of
states (Paper I). In order to understand the numerology
better we consider a generalized Lennard-Jones potential, denoted
LJ(a,b) ($a>b$): 

\begin{equation}
v_{LJ}^{a,b}(r)\propto(\sigma/r)^a - (\sigma/r)^b,
\end{equation}

\nod although the standard 
LJ(12,6) case will be used for most examples. Starting from 
the idea that short distances dominate fluctuations and
that the observed correlations are suggestive of a power-law
interaction, we show that at a given density, the Lennard-Jones
potential may be approximated by a single effective inverse
power law over a range from a little less than $\sigma$ (where $v_{LJ}$ changes
sign) to around $r_m$, the location of the potential minimum 
($r_m=2^{1/6}\sigma$ for LJ(12,6)), covering an energy range of 
approximately $-\epsilon$ to $+2\epsilon$, where $\epsilon$ is the well-depth. 
At first sight,
one might expect that if this were at all the case, the effective power
would be less than $a$, somehow a mixture of the two exponents $a$ and $b$. 
It was noticed by Ben-Amotz and Stell,\cite{Ben-Amotz/Stell:2003}
however, 
that the repulsive core of the LJ(12,6) potential may approximated by an 
inverse power law with an exponent $\sim$18, in agreement with our data
(Paper I). To see how we get an exponent greater than $a$, 
note that the exponent $n$ in an inverse power law can be 
extracted from different ratios of derivatives:

\begin{equation}
  v_{PL}(r) = A r^{-n} +B,
\end{equation}

\nod where $B$ is a constant. This implies that

\begin{align}
n &= \frac{-r v_{PL}'(r)}{(v_{PL}(r)-B)} \nonumber \\
&=  \frac{-r v_{PL}''(r)}{v_{PL}'(r)} -1  \nonumber \\
&= \frac{-r v_{PL}'''(r)}{v_{PL}''(r)}-2 = \ldots
\end{align}


\nod since successive differentiation gives factors of $n$, then $n+1$, etc. 
For a potential $v(r)$ which is not an inverse power law, these expressions
provide different definitions of a local effective power-law exponent
(assuming $v(r) \rightarrow 0$ as $r \rightarrow \infty$):

\begin{align}
n^{{(0)}}(r) &\equiv -r v'(r)/v(r) \nonumber \\
n^{{(1)}}(r) &\equiv -r v''(r)/v'(r) -1  \nonumber \\
n^{{(2)}}(r) &\equiv -r v'''(r)/v''(r) -2 \nonumber \\
n^{{(p)}}(r) &\equiv -r v^{(p+1)}(r)/v^{(p)}(r) -p \label{define_n_p}.
\end{align}

\begin{figure}
\includegraphics[width=3.5 in]{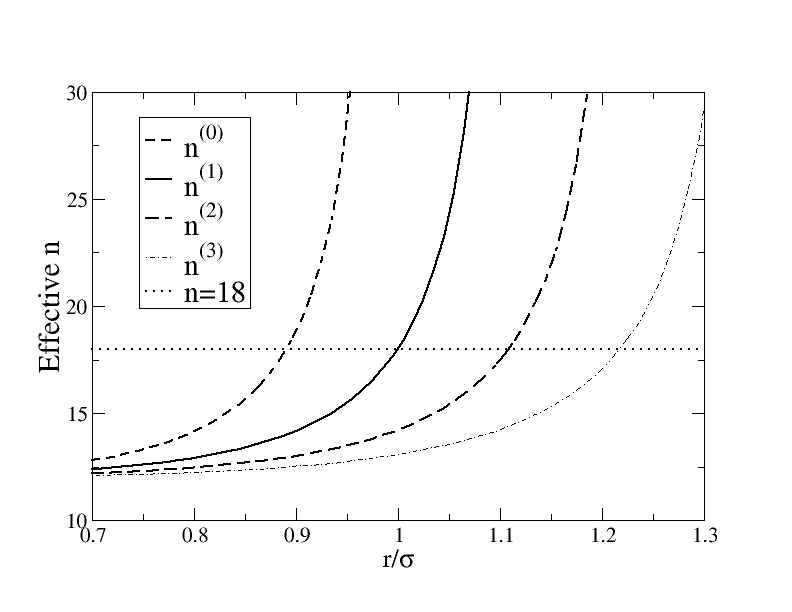}
\caption{\label{differentOrderNs}Effective power-law exponents defined by
 derivative ratios of different orders (Eqs.~(\ref{define_n_p})), for the 
standard Lennard-Jones 
potential LJ(12,6). All converge to 12 at small $r$;
they diverge when the derivative in the denominator vanishes, which happens
for larger $r$, the higher the order of this derivative. The term ``effective
inverse power law'' in this paper refers to a power law chosen to
 match $n^{(1)}$ at some point $r_0< r_m\sim 1.12\sigma$, the potential minimum 
where $n^{(1)}$ diverges. A convenient choice is to match at $r=\sigma$, giving 
18. In subsections \ref{crystal} and \ref{FluctuationModes} we show that 
$n^{(2)}(r)$ plays an important
role in the understanding of fluctuations associated with pair distances close
to the potential minimum ($r_m\sim 1.12\sigma$).}
\end{figure}

\begin{figure}
\includegraphics[width=3.5in]{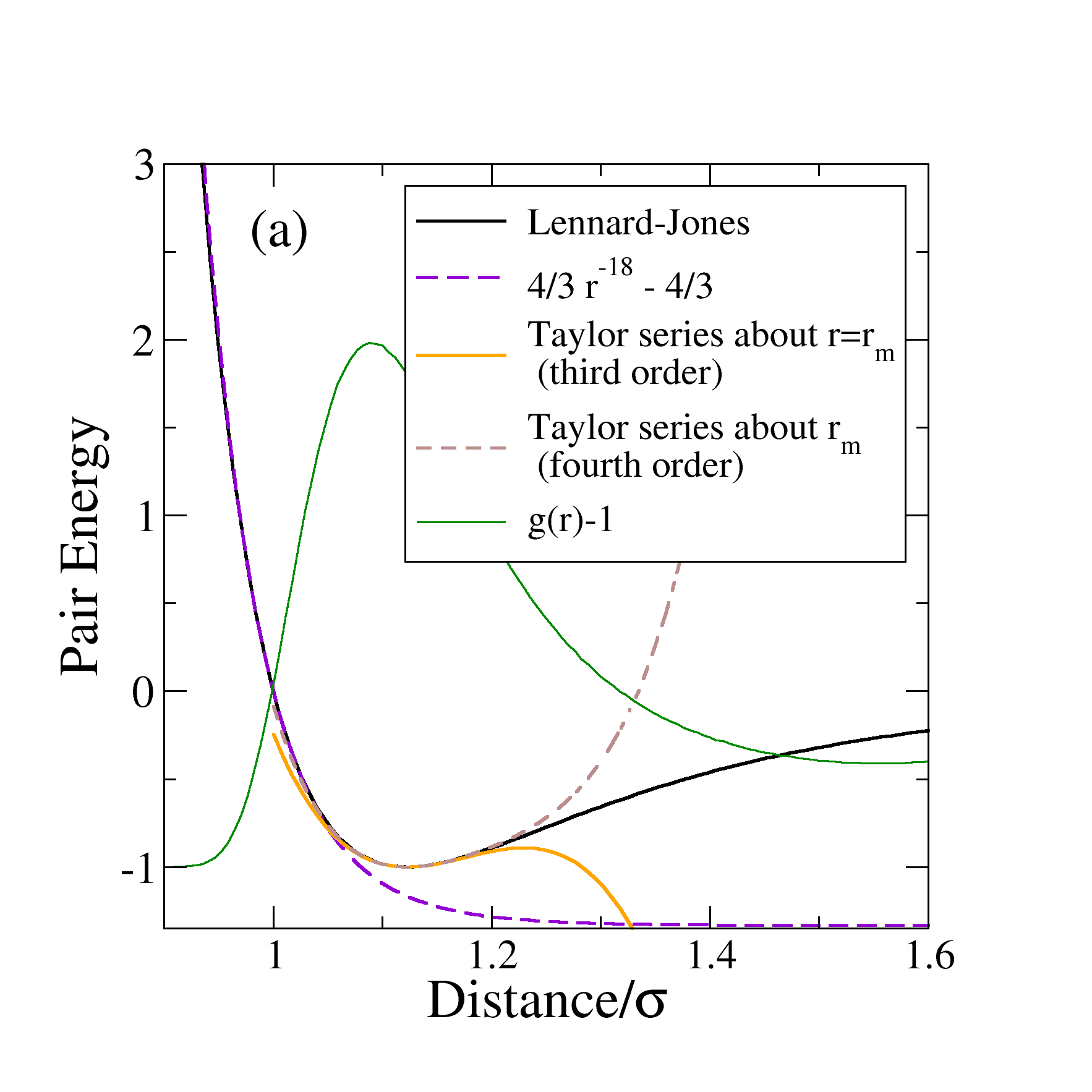} 
\includegraphics[width=3.5in]{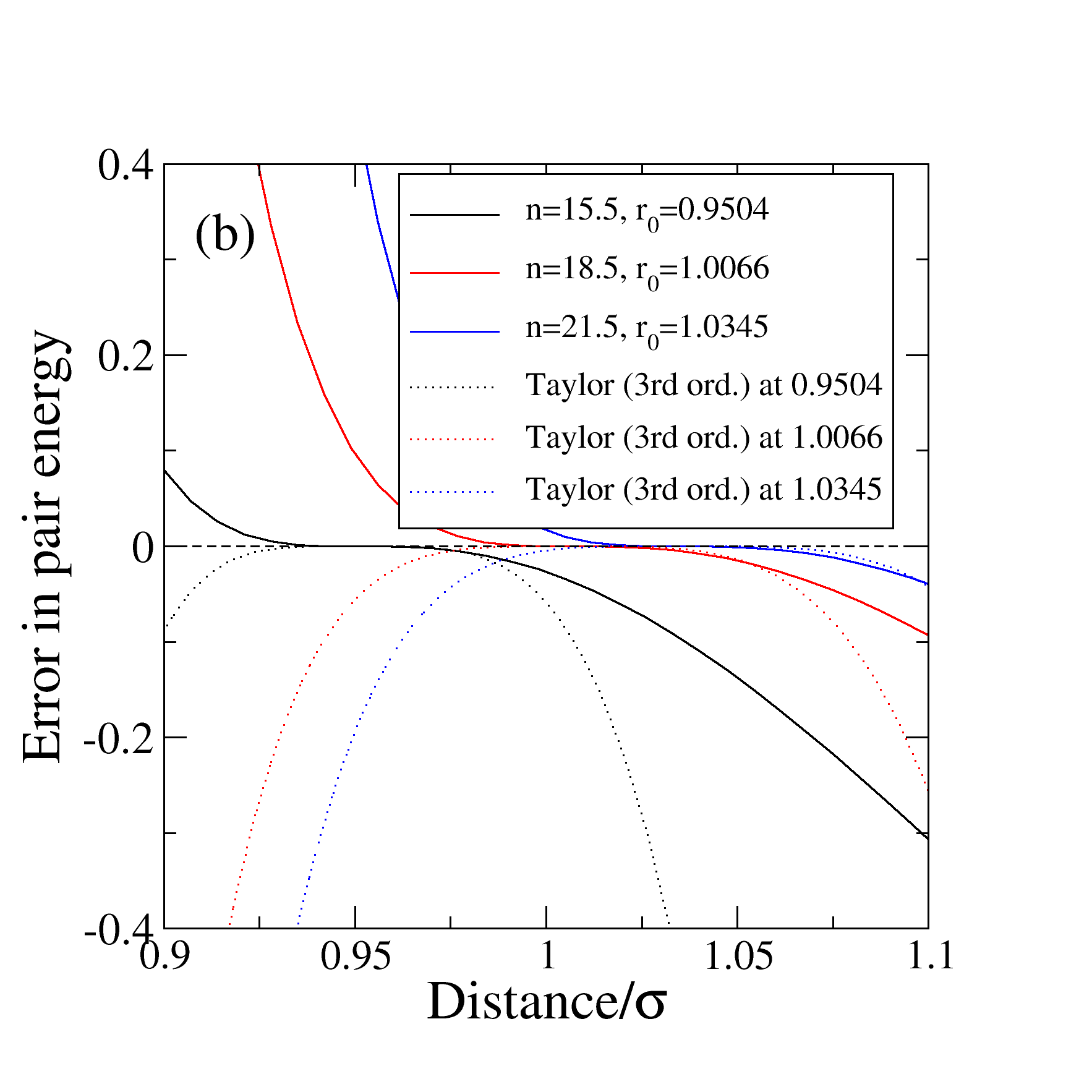} 
\caption{\label{effectivePowerLawFit} (Color online) (a)
  The Lennard-Jones potential $v_{LJ}(r)$ fitted by an  effective power law 
  potential $v_{PL}(r)=Ar^{-n} + B$ covering the most important part of the 
  repulsive part of the potential. The exponent $n$ was chosen to be $18$ which
  optimizes agreement at $r_0=\sigma$, where the effective power law exactly
  matches not just $v_{LJ}$, but also its first two derivatives. Also shown are
  the Taylor series expansions of $v_{LJ}(r)$ about $r=r_m$ up to third and 
  fourth order. The radial distribution function $g(r)-1$ (at $T=80$K, $p=0$) 
  is also shown as a convenient reference for thinking about where 
  contributions to potential energy and virial fluctuations come from. 
  0(b) Error made in approximating
  $v_{LJ}(r)$ with different effective power laws matched at different points
  $r_0$ and with Taylor expansions up to third order about the same point. }
\end{figure}

A plot of the first four of these is shown in
Fig.~\ref{differentOrderNs} for LJ(12,6). All 
converge to $a=12$ at short $r$, as they 
must. All increase with increasing $r$  until the denominator vanishes,
but more slowly, the higher the order $p$. In particular, when $n^{(p)}$ 
diverges it is straightforward 
algebra to show that $n^{(p+1)}$ has the value $a+b+p$. So
$n^{(0)}$ diverges at $r=\sigma$ where the potential is zero, and is 
therefore unsuitable for characterizing the range which we expect to dominate
the fluctuations, from energy $+\epsilon$  to energy $-\epsilon$ (it 
is also sensitive to the presence of an additive constant, unlike
the others). Instead we
use $n^{(1)}$, which at $r=\sigma$ (the zero of $v$ and the divergence of 
$n^{(0)}$ ) takes the value $a+b$, or 18 for the LJ(12,6). Taking the factor 
3 in the denominator of the virial into account, this would explain the 
slope $\sim$6 observed in the simulations (Paper I). But first we should see 
how well an inverse power law with this exponent actually fits the 
Lennard-Jones potential. We denote the point matching point $r_0$. With the 
exponent fixed, we are free to choose the 
multiplicative constant $A$ and the additive one $B$  to match the slope and 
value at $r=\sigma=r_0$; the resulting expression is 
$(4/3)((r/\sigma)^{-18}-1)$. 
This is plotted in Fig.~\ref{effectivePowerLawFit}(a) along with $v_{LJ}$. We 
can match at different values of $r$ by finding the expression for $n^{(1)}(r)$
for the generalized Lennard-Jones potential:

\begin{equation}
n^{(1)}(r) = b + \frac{a-b}{1-(b/a)(r/\sigma)^{a-b}}
\end{equation}

\nod which becomes $6+12/(2-(r/\sigma)^6)$ for 
LJ(12,6).\cite{Pedersen/others:2008}

The fact that we can choose a function (an inverse power law in this case) to 
match a given function
and its first two derivatives is nothing special by itself; after all,
a Taylor series does the same. Examples of matching power laws and
Taylor series up to fourth order, at different values of $r_0$, are shown
in Fig.~\ref{effectivePowerLawFit}(b), where the errors $v_{LJ}(r)-v_{PL}(r)$ or
$v_{LJ}(r)-v_{Taylor}(r)$ are plotted. The magnitude of the errors are
similar in the range of $r$ shown, but note that in the Taylor series it was
necessary to match third derivatives at $r_0$ to achieve this, so the 
inverse power-law description is more compact. Moreover the power-law
 representation
is much more useful when it comes to representing the fluctuations of
total energy and virial, because an inverse power law (and therefore the error)
 flattens out at a
constant value at larger $r$, whereas the polynomial nature of the
Taylor expansion means that away from the point of expansion, the
error diverges rapidly (Fig.~\ref{effectivePowerLawFit}(a)).

\begin{figure}
\includegraphics[width=3.5in]{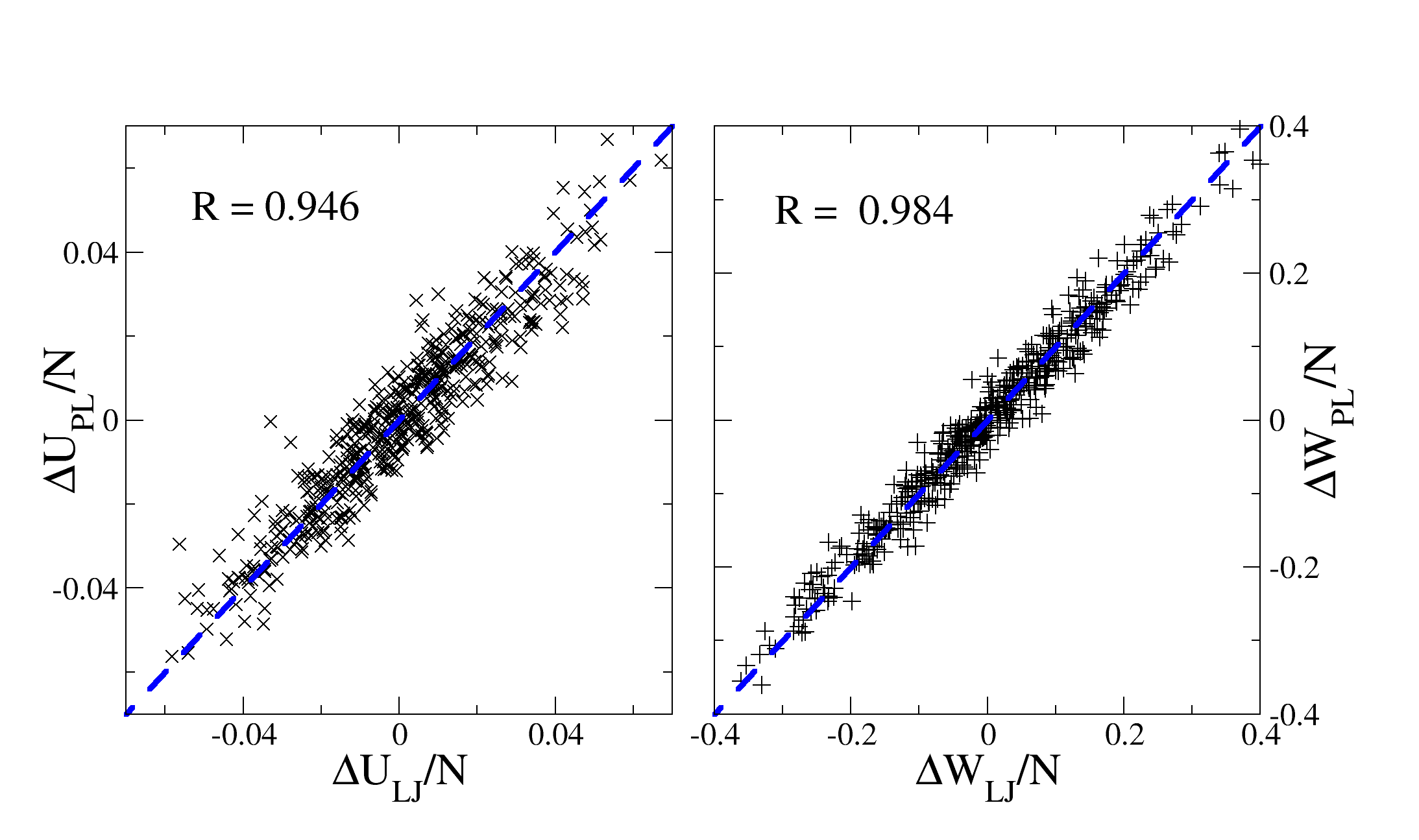}
\caption{\label{effPowerLawWU} (Color online) Scatter-plot of true and 
``reconstructed'' potential-energy and virial fluctuations (dimensionless 
units) for the LJ-liquid, 
where the reconstructed values $U_{PL}$ and $W_{PL}$ were calculated 
from the true configurations, assuming an inverse power-law potential with 
exponent
19.2; mean values have been subtracted off. The state point is the same as in 
Fig.~1 of paper I (zero average pressure, NVT ensemble). The correlation
coefficients are displayed in the figures; the dashed lines indicate
slope unity. The fact that actual 
and reconstructed fluctuations correlate strongly, and with slopes close
to unity, support the idea that the $W,U$ correlation is
derived from an effective inverse power-law potential 
dominating fluctuations.}
\end{figure}


We can test the validity of the power-law approximation for
representing fluctuations in $W$ and $U$ as follows. For the purpose
of computing the energy and virial of a configuration due to a pair
interaction, all necessary information is contained in the
instantaneous radial distribution function (RDF)\cite{Allen/Tildesley:1987}

\begin{equation}\label{timeDepRDF}
g(r,t)  \equiv \frac{2}{N \rho} \sum_{i<j} \delta(r-r_{ij}(t))/(4\pi r^2),
\end{equation}

\nod where $\rho=N/V$ with $N$ and $V$ being the number of particles and the
system volume, respectively. From this $U$ and $W$ may be computed as

\begin{equation}\label{ULJ_from_RDF}
U_{LJ}(t)  = \frac{N}{2} \rho \int_0^{\infty} \! dr \, 4\pi r^2 g(r,t) v_{LJ}(r)
\end{equation}

\begin{equation}\label{WLJ_from_RDF}
W_{LJ}(t)  = -\frac{N}{6} \rho \int_0^{\infty} \! dr \, 4\pi r^2 g(r,t) w_{LJ}(r).
\end{equation}

\nod Now, $U_{LJ}(t)$ is re-written as an inverse power-law potential plus a 
difference term, $U_{LJ}(t) = U_{PL}(t) + U_{\textrm{diff}}(t)$, where

\begin{align}
U_{PL}(t)  &=  \frac{N}{2} \rho \int_0^{\infty} dr\, 4\pi r^2 g(r,t)
v_{PL}(r) \\
U_{\textrm{diff}}(t)  &=  \frac{N}{2} \rho \int_0^{\infty} dr\, 4\pi r^2 g(r,t)
(v_{LJ}(r) - v_{PL}(r))\label{Udiff}
\end{align}



\nod and similarly $W_{LJ}(t) = W_{PL}(t) + W_{\textrm{diff}}(t)$. 
We do not include the additive constant with the power-law
approximation; for practical reasons the potential function should
be close to zero at a cut-off distance $r_c$ (adding a constant to $v_{PL}$
 would only add an overall constant to $U_{PL}$). We refer to $U_{PL}(t)$ and
$W_{PL}(t)$ as ``reconstructed'' potential energy and virial, respectively, to
emphasize that the configurations are drawn from a simulation
using the Lennard-Jones potential, but these quantities are calculated using
the inverse power law. In Fig.~\ref{effPowerLawWU}, 
we show a scatter plot of the true and reconstructed values of $U$ and $W$,
for the same state point ($p$ = 0, $T$ = 80K) as was shown in Figs.~1 and 2(a) 
of Paper I. Here the inverse power-law exponent was chosen to  minimize 
the variances of the difference quantities, $\angleb{(\Delta U_{\textrm{diff}})^2}$
and  $\angleb{(\Delta W_{\textrm{diff}})^2}$. These are minimized for $n=19.3$ and
$19.1$, respectively, so we choose the value 19.2, which corresponds to 
matching the potentials at the distance 1.015$\sigma$.
Note that what are actually plotted are the 
deviations from the respective mean values---the means $\angleb{U_{LJ}}$ and 
$\angleb{U_{PL}}$, for example, do not coincide. But the fluctuations are
clearly highly correlated, and the data lie quite close to the blue dashed
lines, marking slope unity. Specifically, the correlation coefficients are
0.946 for the potential energy and 0.984 for the virial. We can also check how
much of the variance of $U_{LJ}$ is accounted for by $U_{PL}$, 
$\angleb{(\Delta U_{PL})^2}/\angleb{(\Delta U_{LJ})^2}$, and similarly for $W$.
This is a sensible quantity because the \lq PL' and \lq diff' parts are almost 
uncorrelated for the choice $n=19.2$ (cross terms account for 
less than 1\% of the total variance in each case). We find 92\% for $U$ and
95\% for $W$. Thus we see that the power law gives to a quite good 
approximation the fluctuations of $W$ and $U$. The correlation follows from 
this with $\gamma$ given by one third of the effective inverse power-law 
exponent, or 6.4 for this state point. The measured slope 
(Eq.~(\ref{gamma_defined})) was 6.3 
corresponding to an effective exponent 18.9, about 2\% smaller than the 19.3.
A simpler way to determine the exponent would be three times the 
slope, although for some applications it could be advantageous to optimize the
fit as described here.



\subsection{\label{crystal}Low temperature limit: anharmonic vibrations of a crystal}

We turn our attention now
to the fact that the correlation persists even for the crystallized samples
(seen in Paper I in the lower left part of Fig.~4 and in Fig.~6). 
This is not trivial, because the physics of
solids, both crystalline and amorphous systems, is generally dominated by 
fluctuations about mechanically stable structures, and therefore presumably
 (except perhaps at very
 high pressure) by the form of the potential near its minimum $r_m$, i.e.,
including
distances larger than the minimum. Thus the idea
of the effective inverse power law would seem to be inappropriate here---in 
particular since the effective exponent $n^{(1)}$ diverges at $r_m$---and there 
is apparently no reason why one should get a correlation as strong as in the 
liquid and with so similar a slope. In fact there is
an interval of $r$ between $r_m$ and the minimum of the pair virial
$-w(r)/3$ where $v(r)$ is increasing
and $-w(r)/3$ is decreasing, which would lead if anything to a 
negative correlation between $W$ and $U$, when considering individual pair
interactions. Moreover, one would expect that a harmonic approximation of the
potential near the ground-state configuration would be an accurate 
representation of the dynamics in the low-temperature limit, but as we will
see, the harmonic approximation actually implies negative $W,U$ correlation,
which is not observed. In this subsection we show
why the strong positive correlation persists, and why the slope $\gamma$
changes little going from liquid to crystal (at constant volume). 
Although the classical dynamics of a crystal is apparently of little
importance,
since in reality quantum effects dominate, it turns out to be very instructive 
to consider the low-temperature ($T\rightarrow0$) classical limit,
since what we find has significance also in the liquid phase 
(subsection~\ref{FluctuationModes}). The key ideas are (1) that the positive 
correlation emerges only after summing over all interactions---it is
 therefore a collective effect, rather than a single-pair effect, and (2) the
constraint of fixed volume---it is important to recall from Paper I that the 
virial-potential-energy 
correlation only appears under fixed-volume conditions; different volumes give 
approximately the same slope but different offsets (Fig.~4 in Paper I).

\subsubsection{The one-dimensional crystal}

For maximum clarity we start by
considering the simplest possible case, a one-dimensional (1D) crystal with 
periodic boundary conditions and only nearest-neighbor interactions. 
We also suppose that the lattice spacing $a_c$
is equal to the minimum of the potential; this assumption is not made in the
subsequent treatment of the 3D crystal. In a crystal the particles stay
close to their equilibrium positions.  It therefore makes sense to expand the 
pair energy (we have in mind a general pair potential with a single 
minimum) as a Taylor series, leaving out constant terms but keeping
third order terms:

\begin{align}
U &=  \sum_{i} \left(\half k_2(r_{i,i+1}-r_m)^2 + \frac{1}{6}k_3 (r_{i,i+1}-r_m)^3 
\right) \nonumber \\
& \equiv \half k_2 S_2 + \frac{1}{6} k_3 S_3 \label{U_1D}
\end{align}

\nod where $r_{i,i+1}$ is the distance between particles $i$ and 
$i+1$, $k_p$ is the $p$-th derivative of the pair potential at $r=r_m$, and we
introduce the notation 

\begin{equation}
S_p\equiv \sum_i (r_{i,i+1}-r_m)^p.
\end{equation}

\nod The virial is

\begin{align}
W =&  -\frac{1}{3} \sum_{i} \biggl(  k_2 r_{i,i+1} (r_{i,i+1}-r_m)\nonumber \\
& \left.+ \frac{k_3 r_{i,i+1}}{2} (r_{i,i+1}-r_m)^2\right)
\end{align}

\nod which, by writing $r_{i,i+1}=r_m+(r_{i,i+1}-r_m)$, can be rewritten as

\begin{equation}\label{W_1D}
W =  -\frac{1}{3} \left(k_2\ r_m S_1 + k_2 S_2 +\frac{k_3 r_m}{2} S_2
+\frac{k_3}{2} S_3\right). 
\end{equation}

\nod Note that $U$ involves $S_2$ and $S_3$ while $W$ also has a first-order
term with $S_1$. Evaluating the sum $S_1$ is very simple: 
$r_{i,i+1}$
can be expressed in terms of displacements from the equilibrium positions
$u_i$ as $r_{i,i+1}=r_m+u_{i+1}-u_i$, giving for $S_1$

\begin{equation}
S_1 = \sum_i (u_{i+1}-u_i).
\end{equation}

\nod Such a
sum of consecutive relative displacements gives the change in separation
of the two end particles. But the total sum must vanish, because by periodic
boundary conditions the ``end particles'' are the same particle (it doesn't
matter which one), therefore $S_1 = 0$. In fact periodic boundary conditions 
are not necessary, only
that the length is fixed. Since both $U$ and $W$ involve
at lowest order $S_2$, which is positive semi-definite, at 
sufficiently low temperature we may drop the $S_3$ terms. Combining 
 Eqs.~(\ref{U_1D}) and (\ref{W_1D}) we find


\begin{align}
 W &= -\frac{1}{3} \frac{k_2 + k_3r_m/2}{k_2/2} U \nonumber \\
& = \frac{n^{(2)}(r_m)}{3}  U, \label{slope_1D_crystal}
\end{align}

\nod where we have written the coefficient in terms of the $p=2$
effective power-law exponent defined in Eq.~(\ref{define_n_p}).
For LJ(a,b) the coefficient evaluates to $(a+b+1)/3$, which is  
$6.33$ for LJ(12,6), similar to the observed slope.
This short calculation demonstrates the main point: summing over all
 interactions makes the first-order term in the virial vanish, and the 
second-order term is proportional to the second-order term in the potential 
energy. It is also worth noting that for a purely harmonic crystal we can
take $k_3=0$, in which case Eq.~(\ref{slope_1D_crystal}) implies that there is
perfect negative correlation, with a slope of -2/3.

\subsubsection{The three-dimensional crystal}

We now generalize this to three-dimensional crystals, which 
means allowing for transverse displacements. The calculation involves
breaking overall sums into sums over one-dimensional chains within the
 crystal. We also relax
the condition that the lattice constant coincides with the potential minimum,
which is only realistic at low pressures.
 We still assume only nearest-neighbor interactions are relevant (this 
will be justified in the next subsection). Generalization to a disordered 
(amorphous) solid\cite{Binder/Kob:2005} should be possible, since we observe 
the correlation to hold also in that case. The
calculation would necessitate, however, some kind of disorder averaging, which
is beyond the scope of this paper.\footnote{The analysis of fluctuations in 
the liquid in subsection~\ref{FluctuationModes} can be applied to the case of
a disordered solid, explaining the high correlation also there, but it is not
accurate enough to get as good an estimate for the low-temperature limit as
we do our analysis of the crystal.} 

We start by considering a simple cubic (SC) crystal of lattice constant $a_c$,
with interactions only between nearest neighbors, so that the equilibrium 
bond length is $a_c$ for all bonds. The fact that such a crystal is 
mechanically unstable is irrelevant for the calculation. We shall see later 
that the result applies also to, for instance, an FCC crystal. 
We have the same kind of expansions about $r=a_c$
as above for $U$ and $W$, except a linear term is now included, since we no 
longer assume that $a_c=r_m$. An index $b$ is used to represent 
nearest-neighbor bonds and as for the 1D case we define

\begin{equation}
S_p\equiv\sum_b(r_b-a_c)^p.
\end{equation}

\nod We then have for $U$ and $W$

\begin{align}
U &= \sum_{p=1}^\infty \frac{k_p}{p!} \sum_{\textrm{bonds}\, b} (r_b-a_c)^p = 
\sum_{p=1}^\infty \frac{k_p}{p!} S_p \label{Uexpansion3Dcrys} \\
3 W &= \sum_{p=1}^\infty -\frac{k_p}{(p-1)!} 
\sum_{\textrm{bonds}\, b} r_b (r_b-a_c)^{p-1} \nonumber\\
&= -\sum_{p=1}^\infty \frac{k_p a_c }{(p-1)!} S_{p-1} - 
\sum_{p=1}^\infty \frac{k_p}{(p-1)!} S_p \nonumber \\
&= -k_1 a_c S_0 - \sum_{p=1}^\infty \left(\frac{k_p}{(p-1)!} +
 \frac{k_{p+1}a_c}{p!}\right) S_p \label{Wexpansion3Dcrys}
\end{align}

\nod where $k_p$ is the $p$-th derivative of the pair potential at $r=a_c$.
It is convenient to define coefficients $C_p^U$ and $C_p^W$ of the 
dimensionless quantities $S_p/a_c^p$:

\begin{align}
C_p^U & \equiv \frac{k_p a_c^p}{p!}  \label{U_expansion_Cs} \\
C_p^W &\equiv -\left(\frac{k_p a_c^p}{(p-1)!} +
 \frac{k_{p+1}a_c^{p+1}}{p!}\right) \label{W_expansion_Cs}
\end{align}

\nod Dropping the constant term $-k_1 a_c S_0$ in $W$, since it plays no role 
for the fluctuations, we then have 

\begin{align}
  U  &= \sum_{p=1}^\infty C_p^U \frac{S_p}{a_c^p} \\
  3W &= \sum_{p=1}^\infty C_p^W \frac{S_p}{a_c^p} 
\end{align}

\nod The ratio of corresponding coefficients is given by the
$p$th order effective inverse power-law exponent:

\begin{equation}
 C_p^W/ C_p^U =  -\left(p+\frac{k_{p+1} a_c}{k_p}\right)
 = n^{(p)}(a_c).\label{CpW_over_CpU}
\end{equation}

\nod Thus for in the limit of small $a_c$, where these are all similar
and close to the repulsive exponent in the potential 
(Fig.~\ref{differentOrderNs}), the two expansions will 
be proportional to each other, to infinite order. Also worth noting for later
use is that the $C_p$s in each series increase with $p$. For example

\begin{align}
C_2^U / C_1^U &= \frac{k_2a_c}{2k_1}=-\half(n^{(1)}(a_c)+1) 
\label{ratioSuccessiveCoeffsU} \\
C_2^W / C_1^W &= \frac{k_2 a_c+k_3a_c^2/2}{k_1+k_2a_c}
 \nonumber \\
&= -\half (n^{(1)}(a_c)+1) \frac{n^{(2)}(a_c)}{n^{(1)}(a_c)}.
\label{ratioSuccessiveCoeffsW}
\end{align}

\nod For $a_c$ between $\sigma$ and $r_m$, for LJ(12,6), the absolute values of
these ratios lie in the intervals 9.5--$\infty$ and 6.7--9.5 respectively.

From dimensional considerations
the variance of $S_p$ is proportional to $N\sigma_u^{2p}$,
where $\sigma_u^2\propto T$ is the variance of single-particle displacements. At
low $T$, therefore, we expect the $S_1$ terms to dominate, which causes a
 problem since $k_1$ changes sign at $r_m$, corresponding to the divergence of
$n^{(1)}$. In 1D this was avoided by the exact vanishing of $S_1$. In
3D $S_1$ does not vanish exactly, but retains terms second order in
 displacements, and so contributes similarly to $S_2$. It turns out (see below)
 that $S_1$ and $S_2/a_c$ have similar variances and significant
positive correlation, but in view of 
Eqs.~(\ref{ratioSuccessiveCoeffsU}) and (\ref{ratioSuccessiveCoeffsW}) this is not 
even necessary for a strong $W,U$
correlation---the coefficients of the $S_1$ are relatively small so that it is
still the $S_2$ terms that dominate. That is essentially the
explanation of the strong correlations in the crystal, but we now
continue the analysis in more detail in order to
investigate how good it becomes in the limit $T\rightarrow0$. These general
considerations will be of use again in the following subsection,
where we make a similar expansion of the fluctuations in the liquid state.

We need to evaluate $S_1$ and $S_2$ in terms of the relative displacements 
$\bfa{u}_b$ of the pair of particles involved in bond
$b$.\footnote{``Displacement'' refers
 to the displacement of a given particle from its equilibrium position, while 
the ``relative displacement'' is the difference in this quantity for the 
given pair of particles.}  We keep only terms up to second order in 
displacements, since we are interested in the limit of low-temperatures, so
all $S_3$ terms in the expansion are dropped.
In an SC crystal, all nearest neighbor bonds are parallel to one
of the coordinate (crystal) axes. Consider all bonds along the $x$-axis. These
may be grouped into rows of collinear bonds. The sum along
a single row is almost analogous to the one-dimensional case, except
that the bond length $r_b$ now also involves transverse displacements:

\begin{equation}
S_p^{\textrm{row}} =\sum_{b,\textrm{row}} (r_b-a_c)^p
\end{equation}

\nod We write $r_b$ explicitly in terms of relative displacements and expand the
resulting square root, dropping terms of higher order than the second in $u$:

\begin{align}
r_b-a_c &= ((a_c+u_{b,x})^2 + u_{b,y}^2 + u_{b,z}^2)^{1/2}-a_c\nonumber \\ 
&= a_c \left(1+\frac{2u_{b,x}}{a_c} + \frac{u_{b,x}^2}{a_c^2} + \frac{u_{b,y}^2 +
 u_{b,z}^2}{a_c^2}\right)^{1/2}-a_c\nonumber \\ 
&= a_c \left(1+\half\left(\frac{2u_{b,x}}{a_c} + \frac{u_{b,x}^2 + u_{b,y}^2 + u_{b,z}^2}{a_c^2}\right)\right.\nonumber \\ 
& \left. -\frac{1}{8}\left(\frac{2u_{b,x}}{a_c}\right)^2\right) - a_c\nonumber \\
&= u_{b,x}+ \frac{ u_{b,y}^2 + u_{b,z}^2}{2a_c}. \label{rb_m_a}
\end{align}

\nod Now, the sum over bonds in a given row of the parallel displacements $u_{b,x}$
vanishes for the same reasons as in the one-dimensional case. But when
we sum the contributions to $S_1$ over the row, 
there are also second-order terms coming from the transverse displacements.
Extending the sum to all bonds parallel to the $x$-axis, we have part of $S_1$,
denoted $S_1^x$:

\begin{equation}
S_1^{x} =\sum_{b,x} \frac{u_{b,y}^2 + u_{b,z}^2}{2a_c} = \sum_{b,x} 
\frac{|\bfa{u}_{b,\perp}|^2}{2a_c}
\end{equation}

\nod where $\perp$ indicates the component of the relative displacement
vector perpendicular to the bond direction. Writing it this way
allows us to easily include the bonds parallel to the $y$- and $z$-axes,
and the total $S_1$ is given by

\begin{equation}
S_1 =\sum_b \frac{|\bfa{u}_{b,\perp}|^2}{2a_c} \label{S_1_from_u2}.
\end{equation}

\nod Next we calculate $S_2$ to second order in relative
 displacements $\bfa{u}_b$. 
Starting with $S_2^x$, the part containing only
 bonds in the $x$-direction, using Eq.~(\ref{rb_m_a}) we get

\begin{equation}
S_2^{x} =\sum_{b,x} u_{b,x}^2  = \sum_{b,x} |\bfa{u}_{b,\parallel}|^2
\end{equation}

\nod where $\parallel$ means the part of the relative displacement that is 
parallel to the bond. Including all bonds,

\begin{equation}
S_2 =\sum_b |\bfa{u}_{b,\parallel}|^2. \label{S_2_from_u2}
\end{equation}

Now we can return to our expressions for the potential-energy 
fluctuations (Eqs.~(\ref{Uexpansion3Dcrys}) and (\ref{Wexpansion3Dcrys})),
keeping only terms in $S_1$ and $S_2$:

\begin{align}
\Delta U = k_1 S_1 + \frac{k_2}{2} S_2 = 
\frac{k_1}{2a_c}  \sum_b |\bfa{u}_{b,\perp}|^2
+ \frac{k_2}{2} \sum_b |\bfa{u}_{b,\parallel}|^2.
\end{align}

\nod Similarly, for the virial

\begin{align}
3  W &= -k_2 a_c S_1 - \frac{k_3 a_c}{2} S_2 - k_1 S_1 - k_2 S_2 
\nonumber \\
&= -\left(\frac{k_1}{2a_c} + \frac{k_2}{2}\right) \sum_b |\bfa{u}_{b,\perp}|^2
 - \left(k_2+\frac{k_3 a_c}{2}\right) \sum_b |\bfa{u}_{b,\parallel}|^2.
\end{align}

\subsubsection{Statistics of $S_1$ and $S_2$}

It is clear that the $\parallel$ and $\perp$ sums are not equal, 
although they must be correlated
to some extent: If written in terms of single-particle displacements rather
 than relative displacements for bonds, a term such as 
$|\bfa{u}_{b,\parallel}|^2$ for a bond in the $x$-direction becomes 

\begin{equation}
(u_{\bfa{R}+a_c\hat \i,x} - u_{\bfa{R},x})^2 = u_{\bfa{R}+a_c\hat \i,x}^2 + u_{\bfa{R},x}^2 
-2u_{\bfa{R}+a_c\hat \i,x} u_{\bfa{R},x},
\end{equation}

\nod where $\bfa{R}$ is a lattice vector used to index particles. 
Summing over bonds gives

\begin{equation}
S_2 = 2\sum_{\bfa{R}} |\bfa{u}_{\bfa{R}}|^2 -2 \sum \textrm{(para. cross terms)}
\end{equation}

\nod where the cross terms are products of the parallel components of 
displacements on neighboring particles. For $S_1$ we have something similar:

\begin{equation}
S_1 a_c = 2\sum_{\bfa{R}} |\bfa{u}_{\bfa{R}}|^2 -2 \sum \textrm{(perp. cross terms)}
\end{equation}

\nod where here the cross terms involve transverse components. Since the term
$\sum_{\bfa{R}} |\bfa{u}_{\bfa{R}}|^2$ appears in both $S_1$ and $S_2$, these are 
correlated to some extent, but not 100\% since different cross terms appear
(note that if it were 100\%, then $W$ and $U$ would both be proportional
to $S_2\propto S_1$ and also correlated 100\%).
Before considering to what extent they are correlated, let us see 
how much of a difference it makes. Suppose the quantities $S_1 a_c$ and $S_2$ 
have variances $\sigma_1^2$ and $\sigma_2^2$, respectively,
 and are correlated with correlation coefficient $R_S$.
Using the coefficients introduced in Eqs.~(\ref{U_expansion_Cs}) and 
(\ref{W_expansion_Cs})

\begin{align}
 U a_c^2 &= C_1^U (S_1 a_c) + C_2^U S_2 \nonumber \\
3 W a_c^2 &= C_1^W (S_1 a_c) + C_2^W S_2.
\end{align}

\nod From this we
obtain an expression for the $W,U$ correlation coefficient by forming the 
appropriate products and taking averages:

\begin{widetext}

\begin{equation}
R = \frac{C_1^U C_1^W \sigma_1^2 + C_2^U C_2^W \sigma_2^2 
+ (C_1^U C_2^W + C_1^W C_2^U)\sigma_1 \sigma_2 R_S}
{\sqrt{(C_1^U)^2 \sigma_1^2 + (C_2^U)^2 \sigma_2^2 
+ 2C_1^U C_2^U \sigma_1 \sigma_2 R_S} 
\sqrt{(C_1^W)^2 \sigma_1^2 + (C_2^W)^2 \sigma_2^2 
+ 2C_1^W C_2^W \sigma_1 \sigma_2 R_S}} \label{R_from_RS}
\end{equation}

\end{widetext}

\begin{figure}
\includegraphics[width=3.5 in]{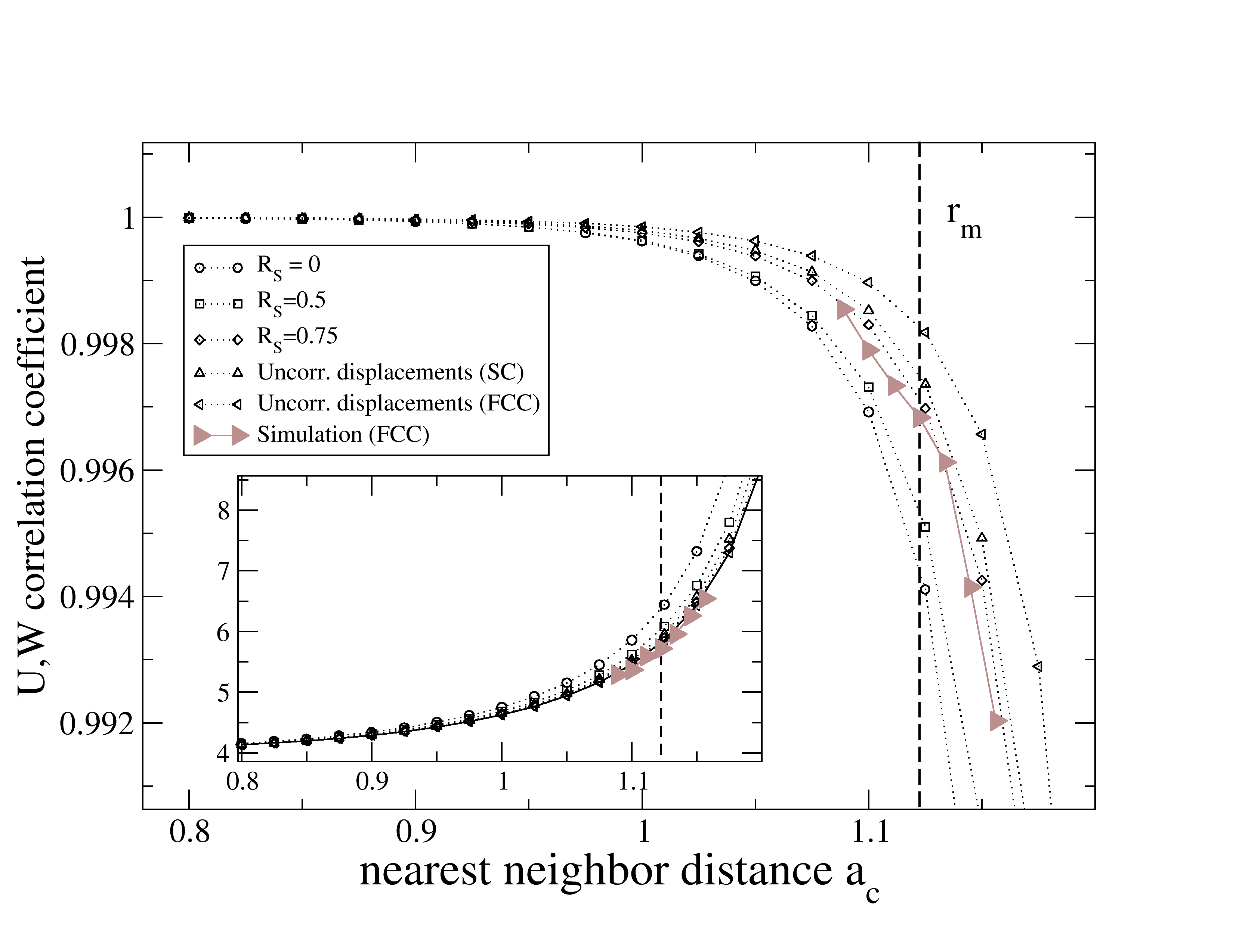}
\caption{\label{S1S2simple} (Color online) Plots of predicted $W,U$ correlation
coefficient for
$T\rightarrow0$ for a crystal of LJ(12,6) particles for different degrees of 
correlation between the quantities $S_1 a_c$ and $S_2$, and of low-temperature
simulation data. The first three curves 
(counting from the bottom) assume the variances of $S_1 a_c$ and $S_2$ 
are equal, and their 
correlation coefficient $R_S$ is 0, 0.5 and 0.75, respectively. The fourth 
curve (up-triangles) results from considering a simple cubic lattice and
assuming individual particles have uncorrelated Gaussian-distributed 
displacements, leading to specific values for the variances and covariance of
$S_1 a_c$ and $S_2$. The fifth (left-triangles) shows the same estimate for an
FCC lattice. The right-triangles are
data from an NVT simulation of a perfect FCC crystal at $T=0.0002$K. 
The conclusion from this figure is that $R$ does not tend to unity as 
$T\rightarrow 0$, although it becomes extremely close.
The inset shows the corresponding slopes $\gamma$ (Eq.~(\ref{gamma_defined})).}
\end{figure}

\begin{table}
\caption{\label{S1_S2_Einstein_stats}Statistics of $S_1 a_c$ and $S_2$ assuming
uncorrelated particle displacements with variance $\sigma_u^2$ for each 
Cartesian component, for simple cubic (SC) and face-centered cubic (FCC) 
lattices.}
\begin{tabular}{|c|c|c|}
\hline
& SC & FCC \\
\hline
$\angleb{S_1 a_c}$ & $6 N\sigma_u^2$ & $12N \sigma_u^2$ \\
$\angleb{S_2}$ & $6 N\sigma_u^2$ & $12N\sigma_u^2$ \\
$\textrm{var}(S_1 a_c)$ & $30 N \sigma_u^4$ & $108N$ \\
$\textrm{var}(S_2$) & $36 N \sigma_u^4$ & $120N$ \\
$\textrm{cov}(S_1 a_c, S_2)$ & $24 N\sigma_u^4$  & $96N$ \\
$R_S$ & 0.73 & 0.84 \\ 
\hline
\end{tabular}
\end{table}

This estimation of $R$ is plotted in Fig.~\ref{S1S2simple} as a function of
lattice constant for $R_S=$0, 0.5 and 0.75, for the case $\sigma_1=\sigma_2$. 
Clearly the value of $R_S$ makes
little difference in the region of interest, $a_c\sim r_m$ or less, where $R$
is above 0.99. Note that
all curves drop dramatically as the lattice constant approaches the inflection
point ($k_2=0$) of the potential (the precise value at which $R$ becomes zero
depends on the statistics of $S_1 a_c$ and $S_2$). In this regime, however, 
higher-order  terms in displacements, including $S_3$, $S_4$, etc., become more 
important, and because of Eq.~(\ref{CpW_over_CpU}) their inclusion tends to
restore $R$ to a high value (we have not calculated their effect in detail). 
Also plotted is the 
estimation of $R$ obtained by assuming that particle displacements are
uncorrelated and Gaussian distributed with variance $\sigma_u^2$ for each
(Cartesian) component, corresponding to an Einstein model of the vibrational 
dynamics.  In this case tedious, but straightforward algebra
allows the means and
(co-)variances of $S_1 a_c$ and $S_2$ to be calculated explicitly for a given
lattice. The results for SC and FCC are given in 
Table~\ref{S1_S2_Einstein_stats}. Notice that the
variance of $S_2$ is somewhat larger than that of $S_1 a_c$, while their means
are equal. This can be traced to the fact the latter contains twice as many
cross terms as the former, and a factor of one half, so 
the contribution from such terms to
the mean is the same in both cases, while the contribution to the variance
is smaller for $S_1 a_c$.  From the (co)-variances we find the correlation
coefficient $R_S=0.73$ and $R_S=0.84$ for the SC and FCC 
cases respectively. These are also plotted
in Fig.~\ref{S1S2simple}. A more exact calculation would 
take into account the true normal modes of the crystal, but would yield 
little of use: 
Data from the crystal simulations at very low temperature, also plotted in
Fig.~\ref{S1S2simple} agree within estimated numerical errors with
both the SC and FCC 
estimates. The key point of this figure---that $R$ is close to but less than 
1---apparently would change little by taking the true crystal 
dynamics into account. In particular it is important to note 
 that if $R=1$ exactly, then this would be true no matter what kind of weighted
average of configurations is taken (what kind of ensemble), so a value less 
than unity in the Einstein
approximation is sufficient to disprove the hypothesis that $R\rightarrow1$ as
$T\rightarrow0$.

\subsubsection{The role of the coefficients $C_{1,2}^{U,W}$}

Since the detailed statistics of $S_1$ and $S_2$ have little effect on the
$W,U$ correlation, it must be mainly due to the numerical values of the
coefficients $C_{1,2}^{U,W}$. We can estimate the effect of these by assuming
$S_1 a_c$ and $S_2$ have equal variance and are uncorrelated ($R_S=0$). Then
according to Eq.~(\ref{R_from_RS}) the $W,U$ correlation coefficient is

\begin{equation}
R = \frac{C_1^U C_1^W + C_2^U C_2^W}
{\sqrt{(C_1^U)^2+ (C_2^U)^2} 
\sqrt{(C_1^W)^2 + (C_2^W)^2} }
\end{equation}

\nod which has the form of the cosine of the angle between two vectors
$\bfa{C}^U\equiv (C_1^U,C_2^U)$ and $\bfa{C}^W\equiv (C_1^W,C_2^W)$. 
Thus the closeness of $R$ to unity indicates that these vectors are nearly
parallel. The tangents of the angles these vectors make with the $C_1$ axis in
$(C_1,C_2)$-space are given by 
Eqs.~(\ref{ratioSuccessiveCoeffsU}) and (\ref{ratioSuccessiveCoeffsW}); 
clearly the two angles become
equal in the limit of small $a_c$, where $n^{(1)}$ and $n^{(2)}$ converge. On the
other hand, for $a_c\sim r_m$ where $k_1 = 0$ and $n^{(1)}$ diverges, the two
vectors are 

\begin{align}
\bfa{C}^U & = (0, k_2/2)a_c^2 \nonumber \\
\bfa{C}^W & = -(k_2,k_2+k_3a_c/2)a_c^2=-k_2 a_c^2(1,1-\half(n^{(2)}+2))\nonumber \\
&= k_2 a_c^2(-1,\frac{n^{(2)}}{2}).
\end{align}

\nod Clearly $\bfa{C}^U$ is parallel to the $C_2$ axis while $\bfa{C}^W$ 
deviates
from it by an angle of order $2/n^{(2)}\sim1/10$. The $W,U$ correlation 
coefficient is then $R=\cos(1/10)\sim 1-\half(1/10)^2\sim 0.995$, in
agreement with the bottom curve (circles) in the main part of 
Fig.~\ref{S1S2simple}. In this case 
($a_c=r_m$, $k_1=0$, $S_1 a_c$ and $S_2$ uncorrelated with equal variance), 
we can obtain a simple expression for the slope

\begin{align}
\gamma &= \sqrt{\frac{(C_1^W)^2 + (C_2^W)^2}{(C_1^U)^2 + (C_2^U)^2}}
= \frac{k_2}{3}\sqrt{\frac{1+(n^{(2)}/2)^2}{(k_2/2)^2}} \nonumber \\
&= \frac{2}{3}\sqrt{1 + (n^{(2)}/2)^2} \sim \frac{n^{(2)}}{3}
\end{align}

\nod consistent with the result from the one-dimensional case. 

Thus when we look at the ``collective'' correlations in the 
crystal we naturally get a slope involving
the effective power-law exponent $n^{(2)}$. Since the latter evaluated at the
potential minimum is similar to $n^{(1)}$ at the zero of the potential, 
the slope is
similar to that seen in the liquid phase. On the other hand, it is more
typical to think about crystal dynamics starting from a harmonic 
approximation, adding in anharmonic terms when necessary for higher
accuracy. How does that work here? If we set $k_3=0$ as well as $k_1=0$, so we
consider the purely harmonic system with the nearest neighbor distance
at the minimum of the potential, then we have $\bfa{C}^U=(0,k_2/2)$ and 
$\bfa{C}^W=-(k_2/3)(1,1)$. These are not close to being parallel, so the 
correlation will be weaker (coming mainly from that of $S_1$ and $S_2$), but 
more particularly, it will be negative, thus qualitatively different from the
anharmonic case. Thus the presence of the $k_3$ 
affects the results at arbitrarily low temperature, so the harmonic
approximation is never good enough. This is reminiscent of thermal
expansion, which does not occur for a purely harmonic crystal. 
In fact, the Gr\"uneisen parameter for a 1D crystal with nearest neighbor 
interactions may be shown\cite{Ashcroft/Mermin:1976} to be equal to $1+n^{(2)}(a_c)/2$.


Finally we consider the more realistic
 FCC crystal. First note that 
Eqs.~(\ref{S_1_from_u2}) and (\ref{S_2_from_u2}) are 
unchanged as long as $a_c$ is now interpreted as the nearest-neighbor
distance rather than the cubic lattice spacing: Each position in an FCC
lattice has 12 nearest neighbors, four located in each of three mutually 
orthogonal planes. Taking the $xy$ and parallel 
planes first, the neighbors are located
along the diagonal directions with respect to the cubic crystal axes. As
 before we can do the sum first over bonds forming a row, then over all 
parallel rows. For a given plane there are two orthogonal sets of rows, but
the form of the sums in Eqs.~(\ref{S_1_from_u2}) and~(\ref{S_2_from_u2})
includes all bonds. The results of the calculation of (co)-variances of
$S_1$ and $S_2$ in the Einstein model of the dynamics are changed in a way that
in fact
increases their mutual correlation and therefore the $W,U$ correlation, as
shown in Table~\ref{S1_S2_Einstein_stats} and Fig.~\ref{S1S2simple}.

To summarize this subsection, the correlation in the crystal is an anharmonic
effect that persists in the limit $T\rightarrow0$. It works because (1) the 
constraint of fixed volume causes the terms in $ U$ and $ W$ that
are first order in particle displacements to cancel
and (2) the coefficients of the
``transverse'' second-order terms are small compared to those of the
``parallel'' ones, a fact which can be traced to the resemblance of the
potential to a power law at distances shorter than the potential minimum. 
``Small'' here means of order 1/10, which leads to over 99\% correlation 
because $R$ is essentially the cosine of this quantity. In one dimension 
there are no transverse displacements and the correlation is 100\% as 
$T\rightarrow0$; in more than one dimension as $T\rightarrow0$ the correlation 
is very close to unity, but never 100\%. 

To gain a more complete
insight into the fluctuations, we next present an analysis which
clarifies exactly the contributions to fluctuations from different
distances, without approximations, now again with the liquid case in mind.

\subsection{\label{FluctuationModes}Fluctuation modes}

In the last two subsections
we considered single-pair effects (associated with $r<r_m$) and collective 
effects (associated with $r\sim r_m$), respectively. In this section we focus on
contributions from particular pair separations without keeping track of
which actual particles are involved.
We identify the contributions to 
$U$ and $W$ coming from all pairs whose separation lies within a fixed small 
interval of separations $r$; fluctuations in the number of such pairs generate
fluctuations in the the contributions. By considering all intervals we can
systematically analyze the variances and covariances of $U$ and $W$ in terms of
pair separation, which is the purpose of this section.

The instantaneous values of $U$ and $W$ are given by Eq.~(\ref{ULJ_from_RDF}) 
and (\ref{WLJ_from_RDF}), generalized to an arbitrary pair potential. By taking
a time (or ensemble) average we get corresponding expressions 
for $\langle U \rangle$ and $\langle W \rangle$ in terms of $g(r) \equiv
\langle g(r,t) \rangle$, the usual thermally averaged RDF. Now we
consider the variances $\langle (\Delta U)^2 \rangle$ and 
$\langle (\Delta W)^2 \rangle$, and the covariance 
$\langle \Delta U\Delta W \rangle$. Starting with, for example, $\Delta U(t) =
 4\pi\rho N/2\int_0^{\infty} dr r^2 \, v(r)\Delta g(r,t)$, where 
$\Delta g(r,t)\equiv g(r,t)-g(r)$, averaging and taking everything except 
$\Delta g(r,t)$ outside the average, we have

\begin{widetext}

\begin{align}
\langle (\Delta U)^2 \rangle &= &  
(4\pi\rho N/2)^2 \int_0^{\infty} \! dr_1 \, r_1^2
\int_0^{\infty} \! dr_2 \, r_2^2  v(r_1) v(r_2) \langle \Delta g(r_1,t) \Delta g(r_2,t) \rangle
\label{UvarIntegral}\\
\langle (\Delta W)^2 \rangle &= & 
(4\pi\rho N/2)^2 \int_0^{\infty} \! dr_1 \, r_1^2
\int_0^{\infty} \! dr_2 \, r_2^2   w(r_1) w(r_2) \langle \Delta g(r_1,t) \Delta g(r_2,t) \rangle 
\label{WvarIntegral} \\
\langle \Delta U \Delta W \rangle & = &  
(4\pi\rho N/2)^2 \int_0^{\infty} \! dr_1 \, r_1^2
\int_0^{\infty} \! dr_2 \, r_2^2  v(r_1) w(r_2) \langle \Delta g(r_1,t) \Delta g(r_2,t) \rangle.
\label{UWcovarIntegral}
\end{align}

\end{widetext}

\nod Clearly the quantity which contains the essential statistical information
about the fluctuations is $\langle\Delta g(r_1,t) \Delta g(r_2,t)
\rangle$, the covariance of the RDF with itself. Its
magnitude is inversely proportional to $N$, so that $\angleb{(\Delta U)^2}$ is
proportional to $N$, as it should be. The variances of $U$ and $W$ and their 
covariance are integrals of this function with different weightings. To make  
further progress, we integrate the integrands for the variances
over $M\times M$ ``blocks'' in $r_1,r_2$-space. This is equivalent to
 considering the energy, say, as the following sum of $M$ 
interval-energies:

\begin{equation}
  U(t) = \sum_{a=1}^M U_a(t) 
\end{equation}

\nod where the interval-energy $U_a(t)$ is defined as the integral between 
boundaries $r_a$ and $r_{a+1}$:

\begin{equation}
  U_a(t) \equiv  \frac{N}{2} \rho \int_{r_a}^{r_{a+1}} dr\, 4\pi r^2 g(r,t) v(r).
\end{equation}

The virial can be similarly represented as a sum of contributions from the 
same $r$-intervals, $W(t) = \sum_{a=1}^M W_a(t)$. From now on we consider 
the primary fluctuating quantities to be $U_a(t)$ and $W_a(t)$ and seek
to understand the correlation between their respective sums in terms of 
correlations between particular $U_a$s and $W_a$s. In order to achieve
 a reasonable degree of spatial resolution, we do not make the intervals 
(blocks) too big, and choose an interval width of 0.05. This gives $M$= 42 
intervals: $U_1$ is the contribution to the energy coming from pairs 
with separation in the range 0.85 to 0.9, marking the lower limit of non-zero
RDF,  while $U_{42}$ refers to the range
2.9 to 2.95, marking the cutoff distance of the potential used here. We shall
see explicitly that only distances up to around 1.4 contribute significantly
to the fluctuations. We denote deviations from the mean as, e.g.,
$\Delta U_a = U_a - \angleb{U_a}$ with the angle brackets representing the
time (or ensemble) average.

\begin{table*}
\caption{\label{covarianceEigenvalues}First ten eigenvalues of the
  super-covariance matrix (Eq.~(\ref{SupCovMtx})), their fractional
contributions to the three (co-)variances 
(Eqs.~(\ref{sumNormalizedDeltaU})--(\ref{sumNormalizedDeltaUW})) and their 
effective slopes (Eq.~(\ref{define_gamma_alpha})), for the SCLJ liquid with 
parameters as in Fig.~1 of Paper I ($\rho$=34.6 mol/l, $T$=80K). Contributions
from the dominant four
eigenvectors are in boldface. The last three rows list sums of the third, 
fourth and fifth columns over, respectively, the dominant four, the
 first ten, and all $2M$ eigenvectors. The sum of the fifth column over all
eigenvectors should be compared (see Eq.~(\ref{sumNormalizedDeltaUW})) to the 
$R=$0.939 listed in Table~I of Paper I.}
\begin{center}
\begin{tabular}{|c|c|c|c|c|c|}
\hline
index & eigenval. & $U$-var. contr. &$W$-var. contr. 
& corr. coeff. contr. & effective slope \\
\hline
1 &  1.73  & 0.01  & 0.01 & 0.01 & 5.38 \\
2 &  1.55  & 0.04  & 0.06 & 0.05 & 7.63 \\
{\bf 3} &  {\bf 1.11}  & {\bf 0.24}  & {\bf 0.15} & {\bf 0.19} & {\bf 4.98} \\
{\bf 4} &  {\bf 0.87}  & {\bf 0.25}  & {\bf 0.25} & {\bf 0.25} & {\bf 6.34} \\
{\bf 5} &  {\bf 0.78}  & {\bf 0.20}  & {\bf 0.14} & {\bf 0.17} & {\bf 5.27} \\
{\bf 6} &  {\bf 0.58}  & {\bf 0.11}  & {\bf 0.17} & {\bf 0.13} & {\bf 7.80} \\
7 &  0.34  & 0.02  & 0.05 & 0.03 & 10.14 \\
8 &  0.23  & 0.01  & 0.03 & 0.01 & 13.67 \\
9 &  0.13  & 0.00  & 0.01 & 0.00 & 116.19 \\
10 &  0.10  & 0.01  & 0.00 & -0.00 & -3.63 \\
\hline
$\sum_{3, \ldots,6}$    & -  & 0.797 & 0.709 & 0.742 & - \\
$\sum_{1, \ldots, 10}$   & -  & 0.884 & 0.877 & 0.849 & - \\
$\sum_{1,\ldots,2M}$ & -  & 1.000 & 1.000 & 0.938 & - \\
\hline
\end{tabular}
\end{center}
\end{table*}

We are interested in the covariance of the $U_a$s with themselves (including
$\angleb{\Delta U_a \Delta U_b}$, $a\neq b$), the
$W_a$s with themselves, and the $U_a$s with the $W_a$s. These covariances
are just what is obtained by integrating the integrands in
Eqs.~(\ref{UvarIntegral}), (\ref{WvarIntegral}), and (\ref{UWcovarIntegral})
over the block 
defined by the corresponding intervals for $r_1$ and $r_2$. These 
values are conveniently represented using matrices $\bfa{\Delta}^{UU}$,
 $\bfa{\Delta}^{WW}$, and $\bfa{\Delta}^{UW}$ defined as 

\begin{equation}
(\bfa{\Delta}^{UU})_{ab}= \langle \Delta U_a \Delta U_b\rangle,
\end{equation} 

\begin{equation}
(\bfa{\Delta}^{WW})_{ab}= \langle \Delta W_a \Delta W_b\rangle,
\end{equation}

\nod and 
\begin{equation}
(\bfa{\Delta}^{UW})_{ab}= \langle \Delta U_a \Delta W_b\rangle.
\end{equation}

\nod Note that the sum of all elements
of $\bfa{\Delta}^{UU}$ is the energy variance, since it reproduces the
 double-integral of Eq.~(\ref{UvarIntegral}); similarly for the other
two matrices:

\begin{align}
\angleb{(\Delta U)^2} & = \sum_{a,b} (\bfa{\Delta}^{UU})_{ab}\\
\angleb{(\Delta W)^2} & = \sum_{a,b} (\bfa{\Delta}^{WW})_{ab}\\
\angleb{\Delta U \Delta W} & = \sum_{a,b} (\bfa{\Delta}^{UW})_{ab}.
\end{align}

At this point we make one further transformation. Define
new matrices $\bfa{\Delta}^{UU*}$, $\bfa{\Delta}^{WW*} $, $\bfa{\Delta}^{UW*}$ by

\begin{equation}
\bfa{\Delta}^{UU*} \equiv \bfa{\Delta}^{UU}/\langle(\Delta U)^2\rangle,
\end{equation}

\begin{equation}
\bfa{\Delta}^{WW*} \equiv \bfa{\Delta}^{WW}/\langle(\Delta W)^2\rangle,
\end{equation}

\nod and

\begin{equation}
\bfa{\Delta}^{UW*} \equiv \bfa{\Delta}^{UW}/ \sqrt{\langle(\Delta U)^2\rangle\langle(\Delta W)^2\rangle}.
\end{equation}

\nod This is equivalent to normalizing the $U_a$s by the standard
deviation $\sqrt{\langle(\Delta U)^2\rangle}$ and the  $W_a$s by
$\sqrt{\langle(\Delta W)^2\rangle}$ respectively. 
The elements of $\bfa{\Delta}^{UU*}$, $\bfa{\Delta}^{WW*}$ and
$\bfa{\Delta}^{UW*}$ can be thought of as representing, in some sense, what
fraction of the total (co-)variance is contributed by the corresponding 
block. 
Normalized in this way, the sum over all elements of $\bfa{\Delta}^{UU*}$ 
and $\bfa{\Delta}^{WW*}$ is exactly unity and that for $\bfa{\Delta}^{UW*}$
equals the correlation coefficient $R$:

\begin{align}
 \sum_{a,b} (\bfa{\Delta}^{UU*})_{ab} &= 1 \label{sumNormalizedDeltaU}\\
 \sum_{a,b} (\bfa{\Delta}^{WW*})_{ab} & = 1 \label{sumNormalizedDeltaW}\\
 \sum_{a,b} (\bfa{\Delta}^{UW*})_{ab} & = R \label{sumNormalizedDeltaUW}.
\end{align}

To make a direct analysis of all possible co-variances, we now
construct a larger  $2M\times2M$ matrix by placing $\bfa{\Delta}^{UU*}$ and 
$\bfa{\Delta}^{WW*}$ on the diagonal blocks, and $\bfa{\Delta}^{UW*}$ and its 
transpose on the off-diagonal blocks:

\begin{equation}
\bfa{\Delta}^{Sup} \equiv \left[\begin{array}{cc}
\bfa{\Delta}^{UU*} & \bfa{\Delta}^{UW*} \\
(\bfa{\Delta}^{UW*})^T & \bfa{\Delta}^{WW*}
\end{array}\right]. \label{SupCovMtx}
\end{equation}

\nod This ``super-covariance'' matrix contains all the information about the 
covariance of the contributions of energy and virial with each other. 
This symmetric and positive semidefinite\footnote{This
  follows since if $v$ is
an arbitrary vector and $A_{ab}=\langle \Delta x_a \Delta x_b\rangle$
is the covariance matrix of a set $\{x_a\}$ of random variables, then 
$v\cdot A \cdot v$ is the variance of the random variable $w=\sum_a v_a x_a$, 
and thus non-negative.} matrix can be separated
into additive contributions by spectral decomposition as

\begin{equation}\label{spectralDecomposition}
\bfa{\Delta}^{Sup} = \sum_\alpha \lambda_\alpha v_\alpha v_\alpha^T
\end{equation}

\nod where
$v_\alpha$ is the normalized eigenvector whose (non-negative) eigenvalue is
$\lambda_\alpha$---this follows from the diagonalization of the matrix. 
Thus we decompose the super-covariance into a sum of parts.
This method of accounting for the variance of many
variables is the basis of the technique known as Principal Component 
Analysis (PCA), which is a workhorse of multivariate data
analysis.\cite{Esbensen/others:2002}  
The eigenvectors represent statistically independent ``modes of
fluctuation''; the corresponding eigenvalue is the part of the
variance within the  2$M$-dimensional space  accounted for by
the mode.  PCA is most effective when the eigenvectors 
associated with the largest few eigenvectors account for most of the 
variance in the set of fluctuating quantities. For example in the extreme 
case where one eigenvector accounts for over 99\% of the variance, we could
claim that all the different apparently random fluctuations of the 
different contributions to energy and virial were moving in a highly
 coordinated way, such that a single parameter (say, the value of any one 
of them) would be enough to give the values of all.

In our case we are not necessarily interested in the modes with the 
largest eigenvalues: a large eigenvalue could describe a fluctuation
mode where the individual 
$U_a$s and $W_a$s change a lot, but their respective sums do not; this would
correspond to the contributions from one interval increasing while those in
others decrease, in such a way that the total is roughly constant. What we
really want are those modes which contribute a lot to the variance of
energy and virial, and to their covariance. This is easy to do by summing all 
elements in the 
appropriate block of the matrix $\lambda_\alpha v_\alpha v_\alpha^T$ where 
$v_\alpha$, $\lambda_\alpha$ are the normalized eigenvector and eigenvalue in
question.\footnote{The sums over the diagonal blocks are non-negative,
  since $\lambda_\alpha$ is, and the sum over the first (last) 
components of $v_\alpha v_\alpha^T$ is the dot product of the first
(last) half of $v_\alpha$ with itself, which is also non-negative.} In 
Table~\ref{covarianceEigenvalues} we list the first ten eigenvalues in 
decreasing order, along with their contributions to the normalized variances 
of energy, virial, and their covariance---the normalized covariance being 
equal to $R_{W.U}$. In addition, an ``effective slope'' for each mode is 
obtained from the $\alpha^{\textrm{th}}$ eigenvector as

\begin{equation}\label{define_gamma_alpha}
\gamma_\alpha = \sqrt{\frac{\langle(\Delta W)^2\rangle}
{\langle(\Delta U)^2\rangle}} \frac{
\sum_{a=M+1}^{2M} (v_\alpha)_a }{\sum_{b=1}^{M} (v_\alpha)_b}
= \gamma \frac{\sum_{a=M+1}^{2M} (v_\alpha)_a }{\sum_{b=1}^{M} (v_\alpha)_b}
\end{equation}

\nod where the numerator gives the sum of virial contributions for that
mode, and the denominator the sum of energy contributions. The factor
in front, which is numerically equal to the overall slope $\gamma$, 
accounts for the standard deviation that we normalized the
 $U_a$s and $W_a$s  by to define the 
matrices $\bfa{\Delta}^{UU*}$,  $\bfa{\Delta}^{WW*}$ and  $\bfa{\Delta}^{UW*}$. 

In the above equation, it looks like $\gamma_\alpha$ is determined by the overall
$\gamma$, whereas we could expect more the opposite, that the overall slope
is somehow an average of the individual effective mode slopes. It looks like 
this because of the normalization choice we made in determining the
 decomposition. We can relate the 
$\gamma_\alpha$ to the $\gamma$ in a more meaningful way by writing the
sums in Eqs.~ (\ref{sumNormalizedDeltaU}) and 
(\ref{sumNormalizedDeltaW}) in terms of the spectral decomposition
Eq.~(\ref{spectralDecomposition}):

\begin{align}
1 &= \frac{\sum_{a,c}\bfa{\Delta}^{WW}_{a,c}}{\sum_{b,d}\bfa{\Delta}^{UU}_{b,d}} = 
\frac{\sum_\alpha \sum_{a,c>M} \lambda_\alpha (v_\alpha)_a (v_\alpha)_c}
{\sum_\beta \sum_{b,d\leq M} \lambda_\beta (v_\beta)_b (v_\beta)_d} \\
&= \frac{\sum_\alpha \lambda_\alpha (\sum_{a > M} (v_\alpha)_a)^2}
{\sum_\beta \lambda_\beta (\sum_{b\leq M} (v_\beta)_b)^2} \\
&= \frac{\sum_\alpha \lambda_\alpha (\gamma_\alpha /\gamma)^2
  (\sum_{a \leq M} (v_\alpha)_a)^2}
{\sum_\beta \lambda_\beta (\sum_{b\leq M} (v_\beta)_b)^2}
\end{align}

\nod where in the last step Eq.~(\ref{define_gamma_alpha}) was used. 
Multiplying both sides by $\gamma^2$ we get an expression for the latter as a
weighted average of the squares of the $\gamma_\alpha$:

\begin{equation}
\gamma^2 = \frac{\sum_\alpha X_\alpha \gamma_\alpha^2 }{\sum_\beta X_\beta}
\end{equation}

\nod where the weight of a given mode slope $\gamma_\alpha$ is (apart from
normalization) $X_\alpha \equiv \lambda_\alpha(\sum_{a \leq M} (v_\alpha)_a)^2 $, 
combining the 
eigenvalue and the square of the summed ``energy part'' of the corresponding
eigenvector.

\begin{figure}
\begin{center}
\includegraphics[width=3.5in]{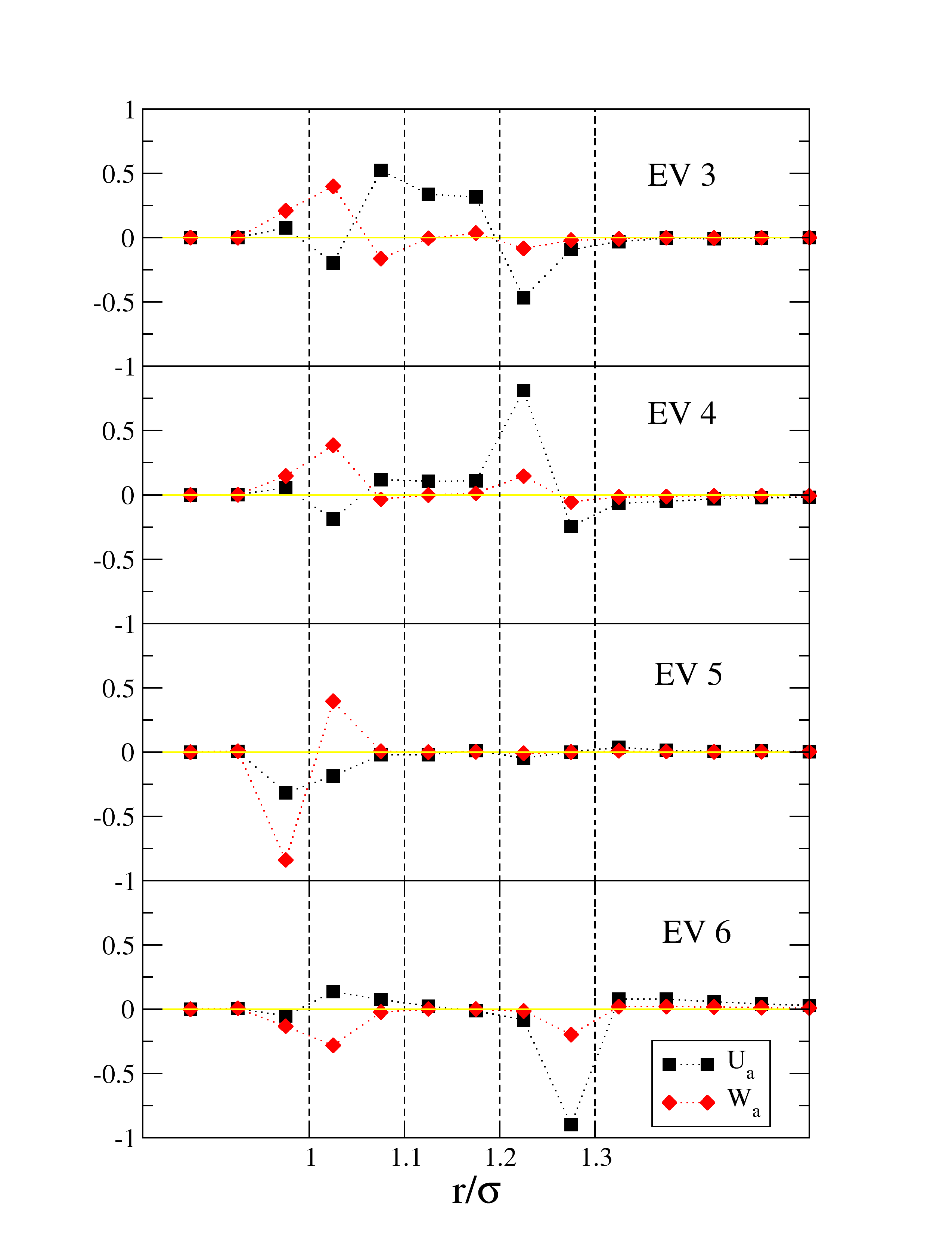}
\end{center}
\caption{\label{fourEigVectors} (Color online)Representations of the 
eigenvectors 3,4,5, and 6 of the super-covariance matrix. Squares 
represent variation in $U_a$ values for a mode; diamonds represent
variation in $W_a$ values.}
\end{figure}

Now we can notice that the third, fourth, fifth and sixth eigenvectors, to be 
referred to respectively as EV3, EV4, EV5, and EV6,
account for most of the three (co-)variances (totalling 0.80 out of
1.00, 0.71 out of 1.00 and 0.74 out of 0.94 for variance of $U$,
variance $W$, and correlation coefficient, respectively). These four
eigenvectors are represented in Fig.~\ref{fourEigVectors}. We observe
that, as expected, most of the fluctuations are
 associated with pair-separations well within the first peak of the RDF, 
which extends to nearly $r=1.6\sigma$ (see 
Fig.~\ref{effectivePowerLawFit}(a)). In fact not much takes place beyond 
$r=1.3\sigma$. Interestingly, of the four, EV5, accounting for less that 20\% 
of the variances, is the only one that directly fits the idea that
the fluctuations take place at short distances, while the other modes extend 
out to $r\sim1.3\sigma$, beyond even the inflection point of the potential
(around 1.24$\sigma$). 

\begin{figure}
\includegraphics[width=3.5 in]{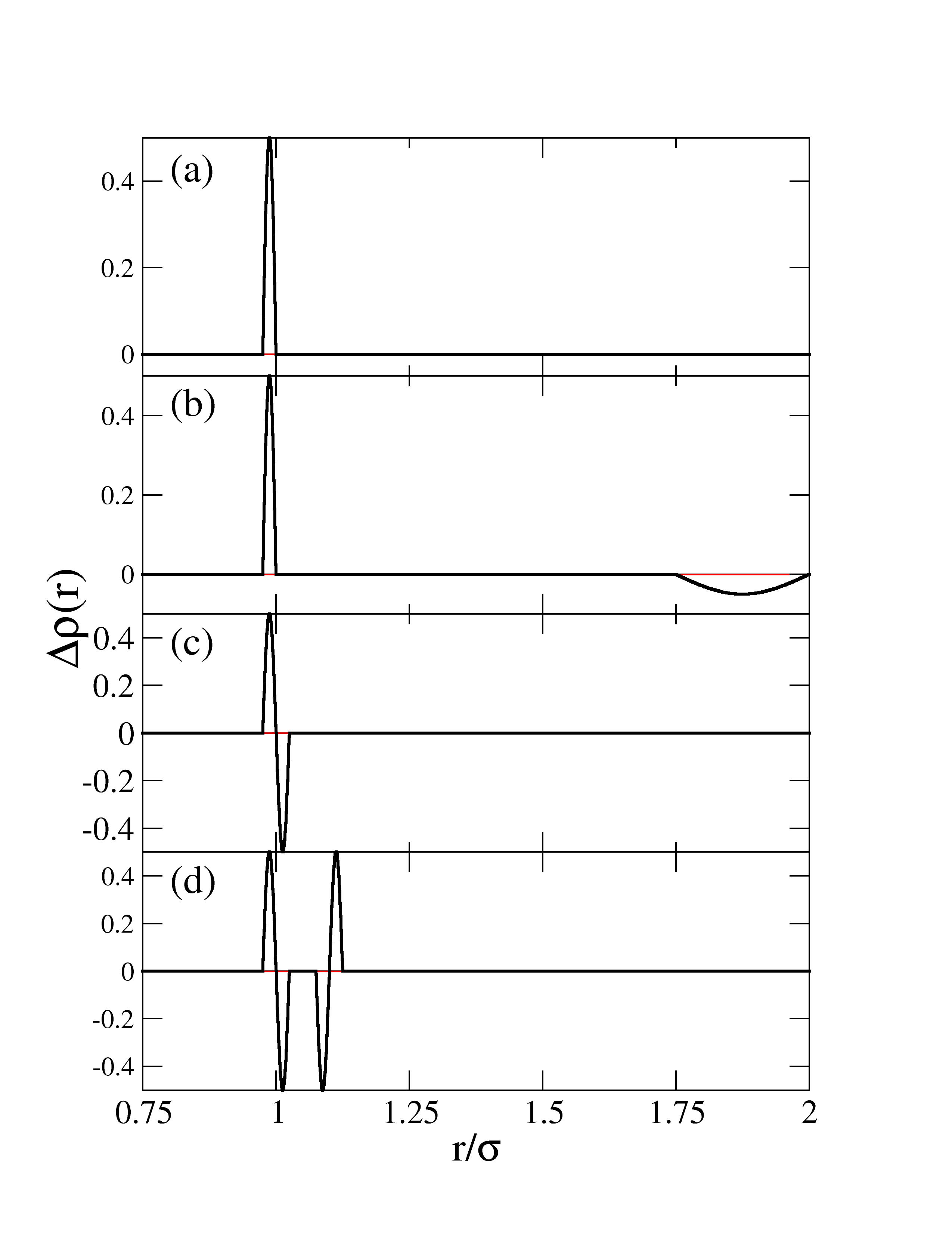}
\caption{\label{illustrateAllowedDeltaNr} Intuitive picture of allowed and
disallowed
fluctuations in $\rho(r)$: (a) is not allowed because it violates the global
constraint $\int \Delta \rho(r)=0$; (b) satisfies the global constraint but not
locality; (c) could correspond, for instance, to a single bond becoming 
shorter, but this is inconsistent with fixed volume (vanishing first 
moment---such a change cannot happen in isolation); (d) is allowed---it 
corresponds, for example, to a single particle being displaced towards one 
neighbor and away from another. Thus one bond shortens and one lengthens.}
\end{figure}

It is instructive to repeat the fluctuation mode analysis for a 
non-strongly correlating liquid, the Dzugutov liquid at $T=0.65$. We do not 
show the full
results here, but they can be summarized as follows. There are two modes which
are concentrated at distances less than and around the first minimum of the
potential. These have slopes of 5.73 and 5.01 and contribute a total of about
0.35--0.4 to the variances and correlation coefficient. Since the latter
 is 0.585 at this temperature, these modes account for most of it. There are 
four
more modes which contribute more than 5\% to the variances, but the slopes are
quite different: -9.34,7.20, 28.43 and -0.67. These four modes all include
significant contributions at distances corresponding to the peak in
$v(r)$; clearly this extra peak in the potential and the associated peak in the
pair-virial $w(r)$ give rise to components in the fluctuations which cannot
be related in the manner of an effective inverse power law, even though 
fluctuations occurring around the minimum can. As a result the overall 
correlation is rather weak.

\subsection{\label{Synthesis}Synthesis: why the effective power-law works even at longer distances}

We can apply ideas similar to those used in the crystal analysis to understand
 why the correlation holds even for modes active 
at separations larger than the minimum, why the slopes are similar to the
effective power-law slopes, and why the effective power law works as well as it
does. Recall the essential ingredients of the crystal analysis:
the importance of summing over all pairs, the fixed-volume constraint and
the increase of magnitude of coefficients of the Taylor expansion with order. 
These
are equally valid here, but now we use them to constrain the allowed 
deviations in $g(r)$ from its equilibrium value, instead of displacements from 
a fixed equilibrium configuration. Define the resolved pair-density $\rho(r)$
by

\begin{equation}\label{rho_of_r}
\rho(r) \equiv (N/2) 4\pi r^2 \rho g(r).
\end{equation}

\nod The requirement that
this integrates to the total number of pairs in the system,
 $\int_0^\infty \rho(r) dr= N(N-1)/2$, gives a global constraint on fluctuations
of $\rho(r)$:

\begin{equation}\label{zerothMomentRho}
\int_0^\infty \Delta \rho(r) dr=0.
\end{equation} 

\nod A typical
fluctuation will have peaks around the peaks of $g(r)$, but only those near the
first peak will significantly affect the potential energy and virial
(subsection \ref{FluctuationModes}). We can 
assume that, for a dense liquid not close to a phase transition, almost any 
configuration $\Gamma$ may be mapped to a nearby reference configuration 
$\Gamma_{\textrm{ref}}$ whose RDF is identical with the thermal
 average $g(r)$. \lq Nearby' implies that the particle displacements relating 
the $\Gamma$ and $\Gamma_{\textrm{ref}}$ are small compared to the inter-particle
spacing.\footnote{ The configuration $\Gamma_{\textrm{ref}}$ is analogous to 
the inherent state configuration often used to describe viscous liquid 
dynamics,\cite{Goldstein:1969,Stillinger:1995} which is obtained
by minimizing the potential energy starting from configuration $\Gamma$.} These
displacements define the deviation $\Delta \rho(r)$ of $\rho(r)$ from its 
equilibrium value. Mapping to $\Gamma_{\textrm{ref}}$ gets around the absence of
a unique equilibrium configuration as in the crystal case. 

Let us consider
what restriction this places on $\Delta \rho(r)$; these are illustrated in 
Fig.~\ref{illustrateAllowedDeltaNr}. Because the displacements are small, 
$\Delta \rho(r)$ must be local: a peak in $\Delta \rho(r)$ at some $r$ 
must be compensated by an opposite peak at a nearby location $r_{\textrm{ref}}$, 
rather than one far away 
(thus example (b) in the figure is not allowed)---this corresponds to a bond
having length $r$ in the actual configuration and length $r_{\textrm{ref}}$ in 
$\Gamma_{\textrm{ref}}$ (Fig.~\ref{illustrateAllowedDeltaNr} (c)).
Finally fixed volume implies that a fluctuation cannot involve
any substantial change in the mean nearest-neighbor
bond length. This may be expressed mathematically as the near vanishing
of the first moment of $\Delta \rho(r)$:

\begin{equation}\label{firstMomentRho}
\int_{\textrm{first peak}} r \Delta \rho(r) dr \cong 0.
\end{equation}

\nod Thus, if a particle is displaced towards a neighbor on one side,
it is displaced away from a neighbor on the opposite side, thus the resulting
fluctuation is expected to 
look like Fig.~\ref{illustrateAllowedDeltaNr}(d), which is 
characterized by having vanishing zeroth and first moments.
Note that we restrict the integral to the first peak. The principle that
fluctuations of $\Delta \rho(r)$ must be local allows us to write a
version of Eq. (\ref{zerothMomentRho}) similarly restricted:

\begin{equation}\label{zerothMomentRhoFirstPeak}
\int_{\textrm{first peak}} \Delta \rho(r) dr = 0
\end{equation}

Equations (\ref{firstMomentRho}) and (\ref{zerothMomentRhoFirstPeak}) cannot 
be literally true, since there must be
 contributions from fluctuations at whatever cut-off distance is used to define
 the boundary of the first peak. For instance, there could be a 
fluctuation such as Fig 6(d) centered just to the right of this cut-off, so 
that only the first positive part was included in the integrals. 
We can ignore these
contributions if we assume that the potential is truncated and shifted to 
zero at the boundary, as is standard in practice (although usually at
 larger distances). Then fluctuations right at the boundary do not contribute
to the potential energy. The fact that the only contributions to the integral
are at the boundary
is a restatement of the locality of fluctuations. \footnote{We have checked 
the statements that the zeroth and first moments of $\rho(r)$ over the first
peak are constant apart from contributions at the cutoff by computing 
orthogonalized versions of them (using Legendre polynomials defined on the 
interval (0.8$\sigma$,1.4$\sigma$)) and showing that they are strongly
 correlated  (correlation coefficient 0.9) with a slope corresponding to the 
cutoff itself.}

Now we make a Taylor series expansion of $v(r)$ around the maximum $r_M$ of the 
first peak of $g(r)$, using $U=\int_0^\infty \rho(r) v(r) dr$,

\begin{align}
\Delta U =&  \int_{\textrm{first peak}} \Delta \rho(r) \Biggl(v(r_M)+ k_1 
(r-r_M)  \nonumber \\
& \left.+\half k_2 (r-r_M)^2 + \ldots\right) \equiv \sum_p \frac{k_p}{p!} M_p.
\end{align}

\nod As for the crystal $k_p$ is the $p$th derivative 
of $v(r)$ at the expansion point ($r_M$ here), while 
$M_p$ is the $p$th moment of $\Delta \rho(r)$:

\begin{equation}
M_p \equiv \int_{\textrm{first peak}} \Delta \rho(r)(r-r_M)^p dr.
\end{equation}

\nod A similar series exists for $W$, with coefficients given by 
Eq.~(\ref{W_expansion_Cs}):

\begin{equation}
3\Delta W =\sum_p \frac{C_p^W}{(r_M)^p} M_p.
\end{equation}

\nod The moments play a role exactly analogous to the sums $S_p$ in the analysis
of the crystal. The near vanishing of $M_1$ corresponds to that of $S_1$ in the
crystal case, both following from the fixed-volume constraint; as there, it
probably holds only to first order in particle displacements (except in one
dimension where it is exact), but we have not tried to make a detailed
 estimate as we did with the crystal. Recalling that the extra contributions to
the $M_2$ terms will be small anyway, in view of 
Eqs.~(\ref{ratioSuccessiveCoeffsU}--\ref{ratioSuccessiveCoeffsW}),
we simply set $M_1=0$, 
so the two series become (noting that $M_0=0$ also)

\begin{align}
\Delta U &= \sum_{p=2}^\infty C_p^U \frac{M_p}{r_M^p} \nonumber \\
3\Delta W &= \sum_{p=2}^\infty C_p^W \frac{M_p}{r_M^p}
\end{align}

\nod where the coefficients $C_p^{U,W}$ are those defined in 
Eqs.~(\ref{U_expansion_Cs}) and (\ref{W_expansion_Cs}), but with $r_M$ 
replacing $a_c$. The points made in subsection~\ref{crystal} regarding the
relation between the two series are equally valid here. At orders $p=2$ and
higher, corresponding coefficients are related by the 
 $n^{(p)}(r_M)$, which are always all above $a=12$.
We expect from dimensional considerations that the variance of $M_p$ is 
proportional to $N w_{FP}^{2p}$, where $w_{FP}\sim 0.3r_M$ is the width of the
first peak of $g(r)$. Thus moments of
higher order should contribute less, and therefore $M_2$ should dominate, 
implying that the proportionality between $\Delta U$ and $\Delta W$ is
 essentially $n^{(2)}(r_M)$. This is in the range 15--24 for $r_M$ in the range
1.05$\sigma$--1.15$\sigma$, giving slopes between 5 and 8, similar to those
observed in the fluctuation modes. Unlike the low-$T$ limit of the
crystal, we cannot assume the fluctuations are particularly localized
around $r_M$, so it is not surprising that a range of slopes show up. Notice 
that we do not see arbitrarily high mode slopes corresponding to the divergence
 of $n^{(2)}(r)$ at the inflection point of the potential. Rather, for modes
centered there, the assumption that we can neglect higher moments of the
fluctuations no longer holds and there is an interpolation between $n^{(2)}(r)$
and $n^{(3)}(r)$ which is smaller (but still greater than $a=12$).

\begin{figure}
\includegraphics[width=3.5 in]{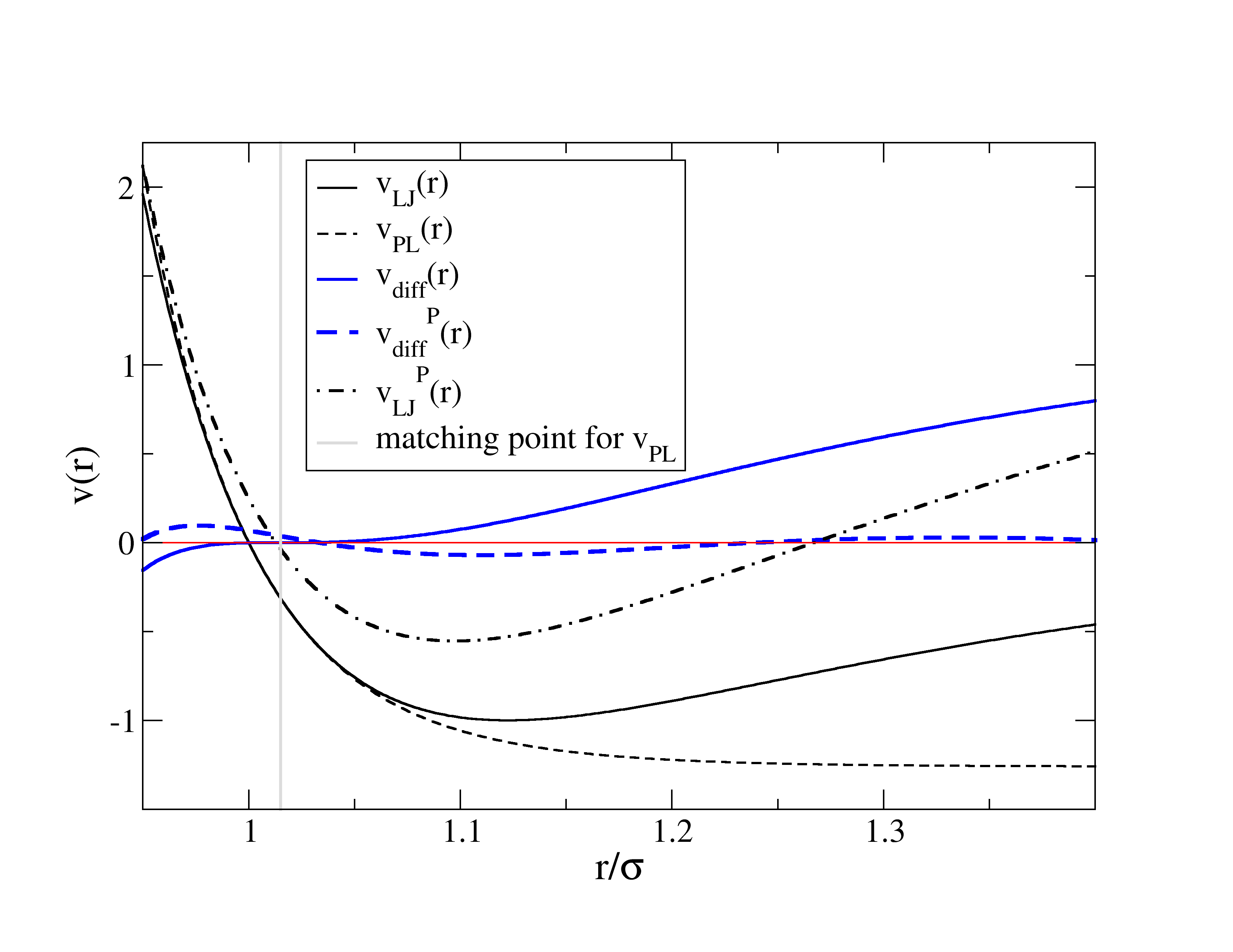}
\caption{\label{whyEffectivePLworks}The true potential $v_{LJ}(r)$, 
the best effective power law $v_{PL}(r)$ (in the sense that the fluctuations in 
potential energy and virial and reproduced most faithfully), and their
difference $v_{\textrm{diff}}(r)$. Also shown are the projected 
versions $v_{LJ}^P(r)$ and
 $v_{\textrm{diff}}^P(r)$ where the constant and linear terms (determined over the
interval $0.95\sigma$ to $1.4\sigma$) have been subtracted off. It is the
projected functions that should be compared in order to make a statement about
the smallness of $v_{\textrm{diff}}(r)$ relative to $v_{LJ}(r)$, since only the
projected functions contribute to fluctuations of total potential energy.}
\end{figure}

We can now understand also why the potential and virial fluctuations, as
``reconstructed'' using the effective power-law potential
 (Fig.~\ref{effPowerLawWU}), agree so well with
the true fluctuations, even though the fluctuation mode analysis shows that
there are significant contributions from distances around and beyond the 
minimum, well away from the matching point $r=1.015\sigma$. 
Fig.~\ref{whyEffectivePLworks} shows the LJ(12,6) potential, the
$n=19.2$ power law (which gave the best fit in 
subsection~\ref{effectivePowerLaw}), and their difference, 
$v_{\textrm{diff}}(r)\equiv v_{LJ}(r)-v_{PL}(r)$.
The latter is obviously very small and flat near the matching
point, but grows significantly in an approximately linear 
fashion at distances larger than $r\sim 1.05\sigma$. In view of
Eqs.~(\ref{Udiff}) and (\ref{rho_of_r}), a fluctuation of $U_{\textrm{diff}}$
can be written as

\begin{equation}
\Delta U_{\textrm{diff}} = \int_0^\infty v_{\textrm{diff}}(r) \Delta \rho(r) dr,
\end{equation}

\nod which has the form of an inner product of functions. Vanishing 
fluctuations of $U_{\textrm{diff}}$ follows if either (1) $v_{\textrm{diff}}$ is 
identically zero, or (2) it is non-zero, but orthogonal to the
space of allowed $\Delta \rho(r)$. Given that $v_{\textrm{diff}}$ is not 
particularly small except close to the matching point
(see Fig.~\ref{whyEffectivePLworks}), the fact that
 $U_{\textrm{diff}}$ fluctuations are relatively small even though they are 
associated with distances away from the matching point indicates that point
(2) must hold approximately. Since allowed $\Delta \rho(r)$ functions are those
with no constant or linear term (see Eqs.~(\ref{zerothMomentRho}) and 
(\ref{firstMomentRho})), functions orthogonal to these are those with only a 
constant and linear term: $f(r)=C f_0(r) + D f_1 (r)$ where $f_0(r)$ is a 
constant function and $f_1(r)$ is a linear function with zero mean over the 
range of interest. It is clear in 
Fig.~\ref{whyEffectivePLworks} that $v_{\textrm{diff}}$ is not exactly of this
form, but it can be well approximated by such a function. This 
approximation can be checked by standard methods of the linear algebra of 
function spaces. First we choose a range $(r_1,r_2)$
over which functions are
to be defined. For purposes this should include the range of significant 
contributions to $W$ and $U$ (Fig.~\ref{fourEigVectors}). We 
choose $r_1=0.95\sigma$ and $r_2=1.4\sigma$. The normalized, mutually 
orthogonal basis vectors $f_0(r)$ and $f_1(r)$ are then given by

\begin{align}
f_0(r) &= 1/\sqrt{r_2-r_1} \nonumber \\
f_1(r) &= \sqrt{\frac{12}{(r_2-r_1)^3}}(r-(r_1+r_2)/2).
\end{align}

The part of $v_{\textrm{diff}}(r)$ which is spanned by these basis functions is
$v_0 f_0 + v_1 f_1$ where $v_i\equiv \int_{r_1}^{r_2} f_i(r) v_{\textrm{diff}}(r) dr$
is the inner product of $v_{\textrm{diff}}(r)$ and the corresponding basis vector.
We define $v_{\textrm{diff}}^P(r)$ as the part of $v_{\textrm{diff}}(r)$ projected
onto the space of allowed functions:

\begin{equation}
v_{\textrm{diff}}^P(r) = v_{\textrm{diff}}(r) - v_0 f_0(r) - v_1 f_1(r).
\end{equation}

\nod This function is also plotted in Fig.~\ref{whyEffectivePLworks}, where it
can be seen that it is certainly small compared to $v_{\textrm{diff}}(r)$ itself.
More importantly, it is also small compared to the projected part of 
$v_{LJ}(r)$, $v_{LJ}^P(r)$, defined analogously, because it is this that explains
why the fluctuations of $U_{\textrm{diff}}$ are small compared to those of $U_{LJ}$
(or equivalently $U_{PL}$). This may be quantified by noting that the ratio of 
their norms is 
$0.09$, which indicates how orthogonal $v_{\textrm{diff}}(r)$ is to the space of
allowed $\Delta\rho(r)$. 
If we projected out only the constant term from $v_{\textrm{diff}}(r)$ and 
$v_{LJ}^P(r)$ (the {\em a priori} more obvious way to
compare the size of two functions) the ratio of norms would be $0.50$, and it
would not be obvious why $v_{PL}$ does as good a job as it does. Thus, again,
constant volume constraint, implying Eq.~(\ref{firstMomentRho}), is important.

The above discussion applies equally well to the virial. We can now write
a more accurate approximate expression for $v_{LJ}(r)$, which we 
call the ``extended effective power-law approximation'':

\begin{equation}\label{v_LJ_approximated}
  v_{LJ}(r) \simeq A r^{-n} + B + C r,
\end{equation}

\nod where $A$, $B$ and $C$ are constants. The associated pair virial 
($w(r)\equiv rv'(r)$) is then

\begin{equation}\label{w_LJ_approximated}
  w_{LJ}(r) \simeq -n A r^{-n} +  C r,
\end{equation}

\nod which has the same form. In both cases the term $Cr$ contributes to
the mean value but not the fluctuations, because $\int r \rho(r) dr$ is 
non-zero, while $\int r \Delta \rho(r) dr=0$ for those $\Delta\rho(r)$ which
are allowed at fixed volume. Note also the contribution to the mean values 
from $Cr$ will depend on volume because $g(r)$, and hence $\rho(r)$, do.
Thus we can see that although there are significant contributions to 
fluctuations away from the matching point where the power law fits the
true potential well, these are essentially equal for both the power-law and the
true potential, because the difference between the two potentials in this
region is almost orthogonal to the allowed fluctuations in $\rho(r)$. This also
explains why the fluctuation only holds at fixed volume (which would not be
explained by the assumption that short-distance encounters dominate the 
fluctuations).

The extended power-law approximation, determined empirically by the 
projection procedure, provides an alternative way to understand
why the effective exponent $n^{(1)}$ evaluated at $r\sim\sigma$ agrees
well with $n^{(2)}$ evaluated around the minimum $r_m$. For the extended
effective power-law approximate, Eq.~(\ref{v_LJ_approximated}) we get

\begin{align}
n^{(1)}(r) &= \frac{-n(n+1)A r^{-(n+1)}}{-n A r^{-(n+1)}+C} - 1\nonumber \\ 
n^{(2)}(r) &= \frac{n(n+1)(n+2)A r^{-(n+2)}}{n(n+1)A r^{-(n+2)}}-2 = n;
\end{align}

\nod Note that $n^{(2)}(r)$ is constant, and  equal to the exponent  $n$ of the 
power-law while $n^{(1)}(r)$ only approaches $n$ when $r$ is small enough that 
the $C$ in the denominator can be neglected (for the true potential 
$n^{(2)}(r)$ increases with $r$ and eventually diverges, see
Fig.~\ref{differentOrderNs}). This emphasizes the greater usefulness
of $n^{(2)}$ in identifying the effective power-law exponent. Recall also
that our analysis earlier in this section indicates that $n^{(2)}(r)$, involving
the second and third derivatives of $v(r)$ near its minimum, is more 
fundamentally the cause of the $W,U$ correlations, explaining something like
80\% of the correlation in the liquid phase and over 99\% in the crystal phase.
The fact that Eq.~(\ref{v_LJ_approximated}) is a good approximation for the 
Lennard-Jones potential pushes the correlation to over 90\% also in the liquid 
phase.

To summarize the last two subsections, we have shown here that the source of 
the 
fluctuations is indeed pair-separations within the first peak, although only
a relatively small fraction of the variances come from the short-$r$ region
where the approximation of the pair potential by a power law is truly valid.
We have also seen how the Taylor-series analysis (which involves the 
crucial step of taking a sum over all pairs) may be extended to cover the whole
first peak area, with all terms giving roughly the same effective slope, given
essentially by the second-order effective exponent: $\gamma \sim 
n^{(2)}(r_m)/3$. The fact that this matches the first-order effective exponent
at the shorter distance $r\sim\sigma$ is equivalent to the extended effective
power-law approximation (Eq.~(\ref{v_LJ_approximated})), which given constant
volume is what justifies the replacement of the potential by
a power law (empirically demonstrated in Fig.~\ref{effPowerLawWU}).

\section{\label{Consequences}Some consequences of strong 
pressure-energy correlations}

This section gives examples of consequences of strong pressure-energy 
correlations. The purpose is show that these are important, whenever present. 
Clearly, more work needs to be done to identify and understand all consequences
of strong pressure energy-correlations.

\subsection{\label{MeasurableConsequences}Measurable consequences of instantaneous $W,U$-correlations}

The observation of strong $W,U$ correlations is of limited interest if
it can only ever be observed in simulations. How can we make a comparison
with experiment? In general, fluctuations of dynamical variables are
related to thermodynamic response 
functions,\cite{Landau/Lifshitz:1980,Hansen/McDonald:1986,Reichl:1998} for 
example those of $U$ are related to the configurational part of the specific 
heat, $C_V^{\textrm{conf}}$. The latter is obtained by subtracting off the 
appropriate kinetic term, which for a monatomic system such as Argon
is $3Nk_B/2$. The virial fluctuations,
however, although related to the bulk modulus, are not directly accessible, 
because of another term that appears in the equation, the so-called 
hypervirial, which is not a thermodynamic quantity.\cite{Allen/Tildesley:1987}
 Fortunately this difficulty can be
handled.

Everything in this section refers to the NVT ensemble. 
First we define the various response functions and configurational
counterparts, the isothermal bulk modulus $K_T$, $C_V$, and the
``pressure coefficient'', $\beta_V$:

\begin{equation}
\begin{array}{rlrl}
  K_T & \equiv -V\left( \frac{\partial \angleb{p}}{\partial V}\right)_T ,&
  K_T^{\textrm{conf}} & \equiv  K_T - \frac{Nk_B T}{V}\\
  C_V & \equiv \left( \frac{\partial E}{\partial T}\right)_V , &
  C_V^{\textrm{conf}} & \equiv  C_V - \frac{3}{2}N k_B  \\
  \beta_V & \equiv \left( \frac{\partial \angleb{p}}{\partial T}\right)_V , &
  \beta_V^{\textrm{conf}} & \equiv \beta_V - \frac{N k_B}{V} \\
  & &   p^{\textrm{conf}} & \equiv p - \frac{N k_B T}{V} = \frac{W}{V}. 
\end{array}
\end{equation}

\nod We also define $c_V\equiv C_V/V$. The following fluctuation 
formulas are standard (see for example Ref.~\onlinecite{Allen/Tildesley:1987})

\begin{align}
\frac{\langle (\Delta W)^2\rangle}{k_B T V} &= \frac{N k_BT}{V} + 
\frac{\angleb {W}}{V}  -K_T + \frac{ \angleb{X} }{V} \label{K_T_fluctForm}\\
\frac{\langle (\Delta U)^2\rangle}{k_B T^2} &= C_V - \frac{3}{2}N k_B 
= C_V^{\textrm{conf}} \\
\frac{\langle \Delta U \Delta W\rangle}{k_B T^2} &= V \beta_V - Nk_B = 
V\beta_V^{\textrm{conf}}.  \label{betaV_fluctForm}
\end{align}

\nod Here $X$ is the so-called ``hypervirial'', which
gives the change of virial upon an instantaneous volumetric scaling of 
positions. It is not a thermodynamic quantity and cannot be determined 
experimentally, although it is easy to compute in simulations. For a pair 
potential $v(r)$, $X$ is a pair-sum,

\begin{equation}
X=\sum_{\textrm{pairs}} x(r)/9
\end{equation}

\nod where $x(r)=r w'(r)$. If we use the extended effective power-law 
approximation (including the linear term) discussed in the last section, then 
from Eq.~(\ref{w_LJ_approximated}) we get $x(r)\simeq n^2  A r^{-n} + Cr$. 
Summing over all pairs, and recalling that when the volume is fixed the $Cr$ 
term gives a constant, we have a relation between the total virial and total
hypervirial, 

\begin{equation}
 X = (n/3) W + \textrm{constant}. \label{X_nover3_W}
\end{equation}

This constant survives, of course, when we take the thermal average 
$\angleb{X}$, as do the corresponding constants in $\angleb{U}$,
 $\angleb{W}$. To get rid of these constants one possibility
would be to take derivatives with respect to $T$, but this can be
problematic when analyzing experimental data. Instead we
simply compare quantities at any temperature to those at some
reference temperature $T_{\textrm{ref}}$; this effectively subtracts off the 
unknown
constants. Taking first the square of the correlation coefficient, we have

\begin{equation}
R^2 = \frac{(\langle \Delta U \Delta W\rangle)^2}
{\langle (\Delta U)^2\rangle\langle (\Delta W)^2\rangle},
\end{equation}

\nod which implies

\begin{equation}
R^2\frac{\langle (\Delta W)^2\rangle}{k_B TV} =
\frac{1}{k_B TV} \frac{(\langle \Delta U \Delta W\rangle)^2}
{\langle (\Delta U)^2\rangle}.
\end{equation}

\nod Inserting the fluctuation formulas Eqs.~(\ref{K_T_fluctForm}) and
(\ref{betaV_fluctForm}) gives 

\begin{align}
R^2 ( \angleb{p} - K_T + \frac{\angleb{X}}{V}) = & 
\frac{1}{k_B TV}\frac{(k_BT^2 V\beta_V^{\textrm{conf}})^2}{k_BT^2 C_V^{\textrm{conf}}} \\
& = T\frac{(\beta_V^{\textrm{conf}})^2}{c_V^{\textrm{conf}}}.
\end{align}

\nod Defining quantities $\tilde{A}\equiv \angleb{p} - K_T + 
\angleb{X}/V$ and $B\equiv  
T(\beta_V^{\textrm{conf}})^2/c_V^{\textrm{conf}}$ (the reason for the tilde on 
$A$ will become clear), we have $R^2 \tilde{A}
= B$. This is an exact relation. To deal with the
hypervirial we first take differences with the quantities at $T_{\textrm{ref}}$,
assuming that the variation of $R$ is much smaller than the $\tilde{A}$
and $B$ variations:

\begin{equation}
R^2 (\tilde{A}-\tilde{A}_{\textrm{ref}}) = B-B_{\textrm{ref}}
\end{equation}

\nod where $\tilde{A}_{\textrm{ref}}=\tilde{A}(T_{\textrm{ref}})$, etc. 
$\tilde{A}-\tilde{A}_{\textrm{ref}}$ written out explicitly is 

\begin{equation}\label{A_Aref}
\tilde{A}-\tilde{A}_{\textrm{ref}} = (\angleb{p}-K_T) - 
(\angleb{p}_{\textrm{ref}}-{K_T}_{\textrm{ref}}) + 
\frac{ \angleb{X} - \angleb{X}_{\textrm{ref}} }{V}
\end{equation}

\begin{figure}
\includegraphics[width=3.5in]{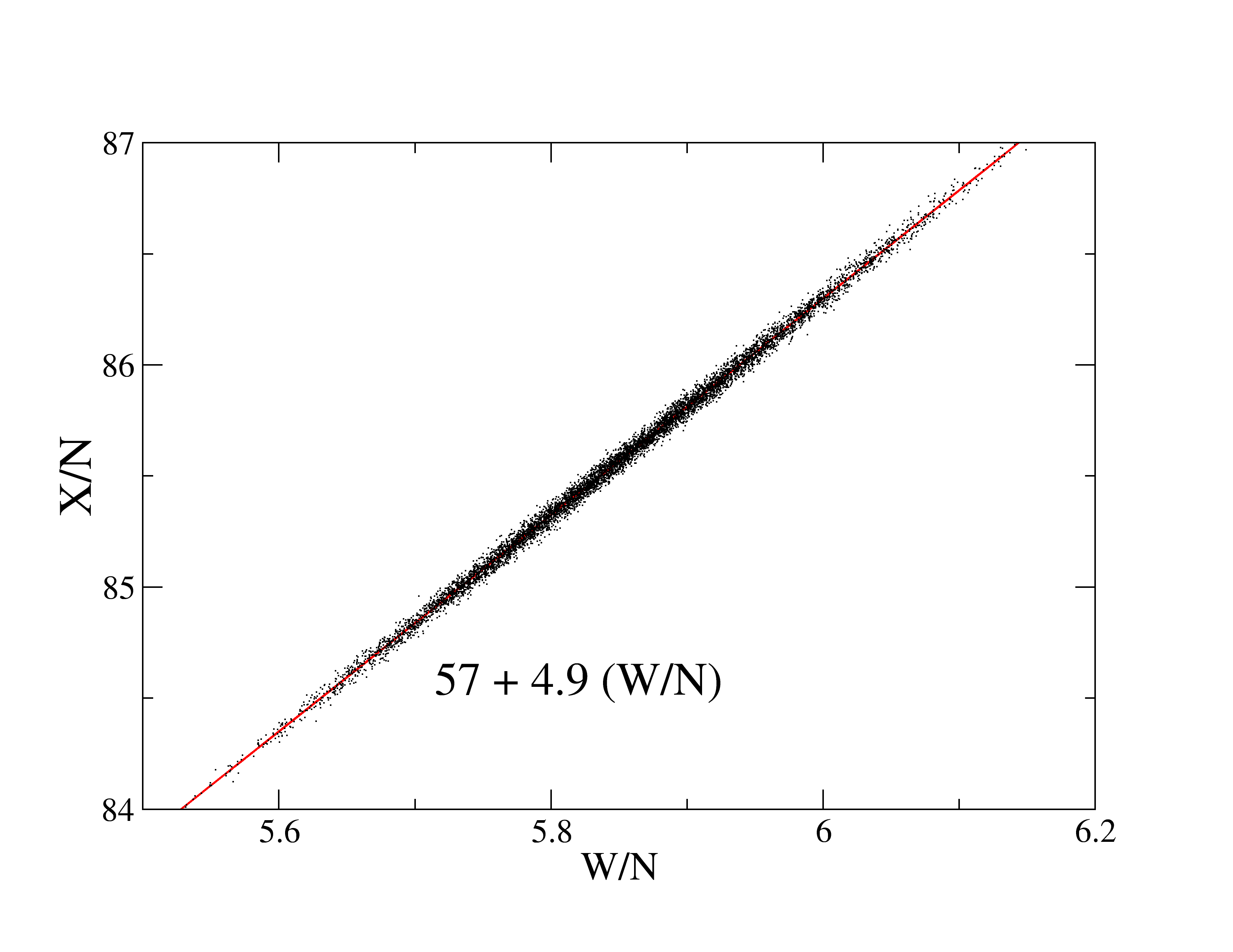}
\caption{\label{VirialHyperVirialArgonT0.80} Scatter-plot of instantaneous 
virial and hypervirial (in dimensionless units) for a SCLJ system at 
$\rho=1.0,T=0.80$ (NVE). The correlation coefficient between these quantities 
is  0.998. The hypervirial is the main contribution to the configurational part
of the bulk modulus; it gives (after dividing by volume) the change of virial 
for a given relative change in volume. The sizeable
constant term in the linear fit shows that Eq.~(\ref{X_nover3_W}) 
is a poor approximation. The slope is 4.9, about 10\% smaller than 
$\gamma\sim5.4$ for this state point. The difference reflects the limit of the
validity of the power-law description--in fact a more detailed analysis shows
that the relation between $W$ and $X$ is dominated by $n^{(3)}(r)$, which is 
smaller than $n^{(2)}(r)$ (Fig.~\ref{differentOrderNs}).}
\end{figure}

\nod Next we use the power-law approximation to replace $ \angleb{X} 
- \angleb{X_{\textrm{ref}}}$ with $(n/3)(\angleb{W}-\angleb{W_{\textrm{ref}}})$. 
This is the crucial point: Whereas it is often not a good approximation that 
$\angleb{X}=(n/3)\angleb{W}$ due to the unknown additive constants
 discussed above, subtracting two state points considers {\it
   changes} of $\angleb{X}$ and $\angleb{W}$ with temperature. Recall from 
section III B of Paper I
that the changes in mean-values $\Delta \angleb{W}$ and $\Delta \angleb{X}$ 
between (nearby)
temperatures are related as the linear regression of the fluctuations of those
quantities at one (nearby or intermediate) temperature. The linear relation
between subtracted mean values holds if the instantaneous $W$ and $X$ are 
strongly 
correlated in the region of interest. The latter is confirmed by our
simulations; indeed the correlation of instantaneous values of $X$ and $W$ is
even stronger than for $W$ and $U$, with approximately the same slope 
(Fig.~\ref{VirialHyperVirialArgonT0.80}). Thus Eq.~(\ref{A_Aref}) becomes

\begin{align}
\tilde{A}-\tilde{A}_{\textrm{ref}} &\simeq (\angleb{p}-K_T) - 
(\angleb{p}_{\textrm{ref}}-{K_T}_{\textrm{ref}}) + \frac{n}{3}\frac{ \angleb{W} -
 \angleb{W}_{\textrm{ref}} }{V} \\ 
& = A-A_{\textrm{ref}}
\end{align}

\nod where $A\equiv p - K_T + (n/3)(\angleb{W}/V)$ (no tilde) contains 
quantities which are all directly
accessible to experiment, except for the effective power-law exponent
$n$. This can be obtained by noting  that if there were perfect correlation,
one could 
interchange $\Delta W$ and $(n/3)\Delta U$; thus

\begin{equation} \label{determiningSlope}
 \frac{\beta_V^{\textrm{conf}}}{C_V^{\textrm{conf}}/V} =  
\frac{\beta_V^{\textrm{conf}}}{c_V^{\textrm{conf}}} =\frac{\langle \Delta U \Delta W\rangle}
{\langle (\Delta U)^2\rangle} = \frac{n}{3}
\end{equation}

\nod which gives for $A$

\begin{equation}
A = \angleb{p} - K_T  + \frac{\angleb{p}^{\textrm{conf}}\beta_V^{\textrm{conf}}}{c_V^{\textrm{conf}}}.
\end{equation}

\begin{figure}
\includegraphics[width=3.5in]{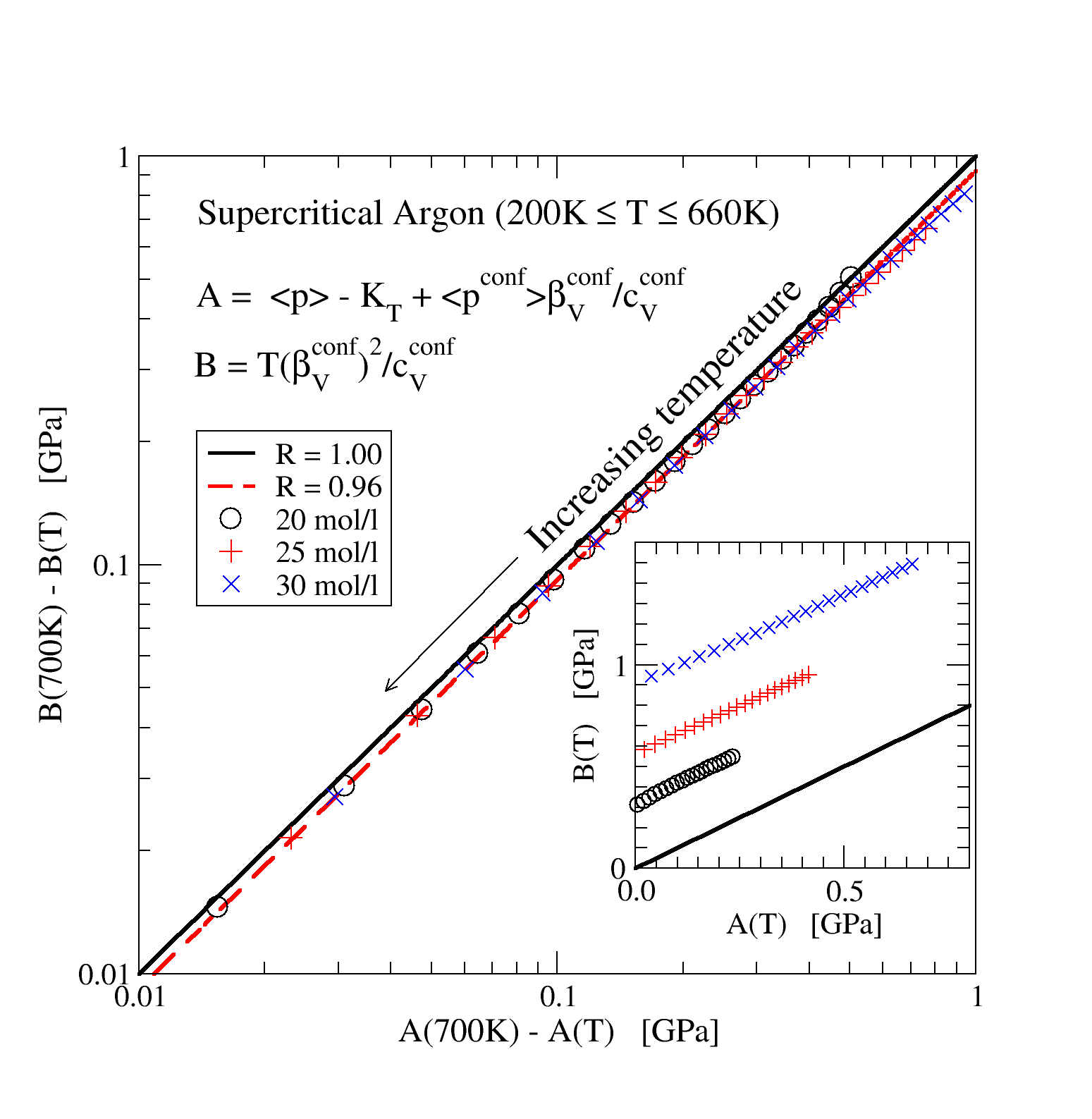}
\caption{\label{ArgonData_xy_log_inset_6} (Color online) Data from the NIST
database\cite{Lemmon/McLinden/Friend:2005} for supercritical
Argon at three different densities covering the temperature range 
200K-660K showing a strong
virial-potential energy correlation ($R=0.96$)
[reproduced from Ref.~\onlinecite{Pedersen/others:2008}]. 
Here $K_T\equiv-V(\partial
\angleb{p}/\partial V)_T$, $p^{\textrm{conf}}\equiv p-Nk_B T/V = W/V$
$\beta_V^{\textrm{conf}} \equiv (\partial 
\angleb{p}/\partial T)_V - Nk_B/V$, and 
$c_V^{\textrm{conf}}\equiv C_V/V-(3/2)Nk_B/V$
The diagonal line corresponds to perfect correlation. The inset shows 
``unsubtracted'' values for $A$ and $B$; the fact that the data do not fall
on the solid line indicates that a power-law description does not hold for the
full thermodynamics.}
\end{figure}

Thus to compare with experiment one should plot
$B-B_{\textrm{ref}}$ against $A-A_{\textrm{ref}}$; the prediction, in
the case of near perfect correlation, $R^2\simeq1$, is
a straight line with slope close to
unity. Fig.~\ref{ArgonData_xy_log_inset_6} shows data for Argon for $T$ between
200K and 660K. Argon was chosen because as a monatomic system there are no 
rotational or vibrational modes contributing to the heat capacity and it is
 therefore straightforward to subtract off the kinetic part.
Also we restrict to relatively high temperature
 to avoid quantum effects. The correlation coefficient $R$ given as the 
square root of the slope of a linear fit
 is 0.96. Note that the assumed constancy of $R$
 is confirmed (going to lower temperatures there are
increasingly large deviations). The importance of subtracting from a reference
state point is highlighted by the inset, which shows that $A(T)=B(T)$ does not
hold: There is a correlation in the fluctuations which is not present in the
full equation of state.\cite{Pedersen/others:2008}

\subsection{\label{ViscousLiquids}Time averaging: Pressure-energy correlations
 in highly viscous liquids}

We have observed and discussed in paper I that when volume is held constant, 
the
correlations tend to become more perfect with increasing $T$, while along an 
isobar (considering still fixed-volume simulations, choosing the volume  
to give a prescribed average pressure), they become more perfect with 
decreasing $T$. This fact makes the presence of correlations highly relevant 
for the physics of highly viscous liquids approaching the glass transition. 
Basic questions such as the origins of non-exponential 
relaxations and non-Arrhenius temperature dependence are still vigorously
debated in this field of 
research.\cite{Kauzmann:1948,Brawer:1985,Angell/others:2000,Dyre:2006a} 
Instantaneous correlations of the kind discussed in this work would seem to
be relevant only to the high frequency properties of a highly viscous
liquid; their relevance to the long time scales on which structural
relaxation occurs follows from the separation of time scales as explained
below.

\begin{figure}
\includegraphics[width=3.5 in]{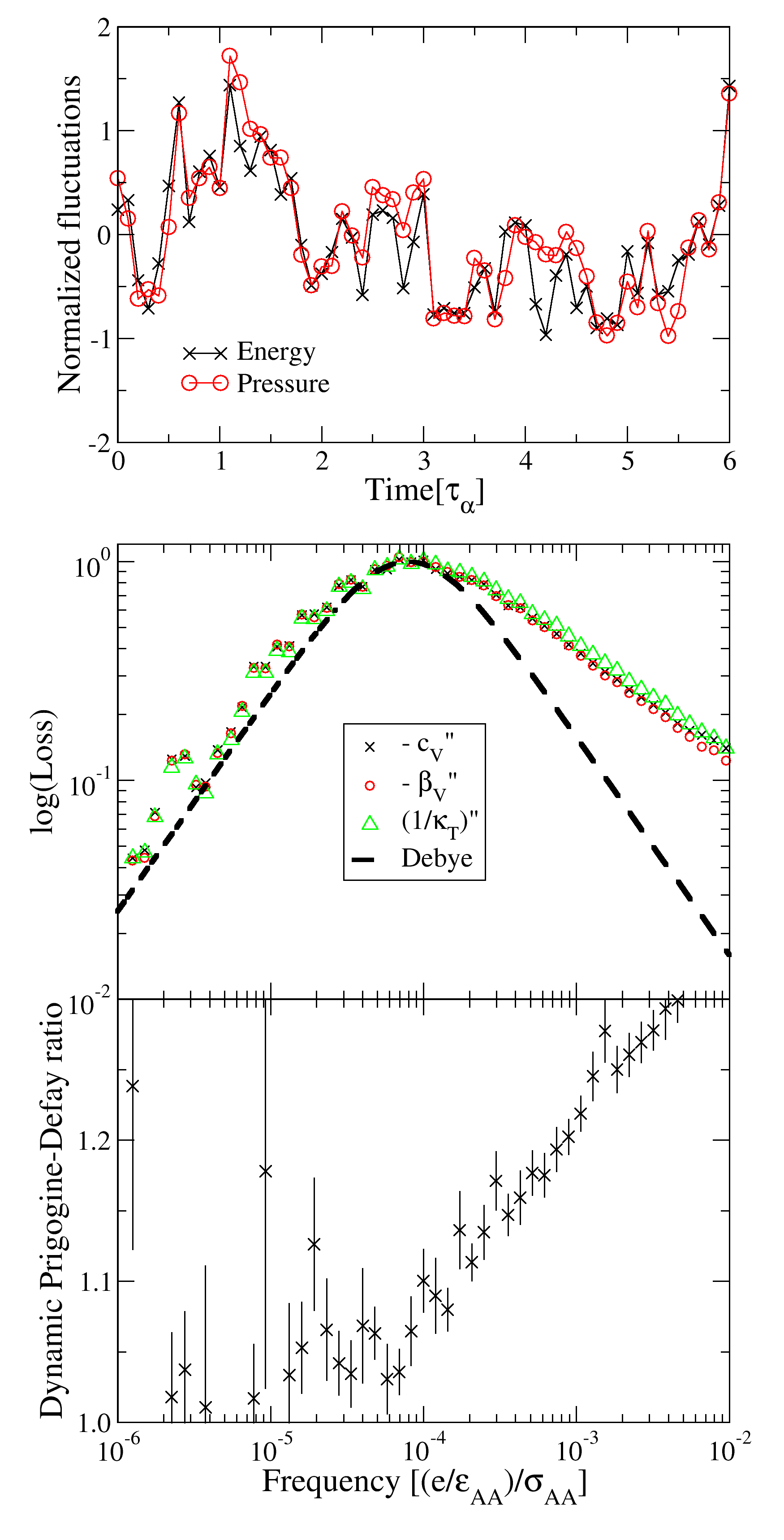}
\caption{\label{combinedPDR} (Color online)
  Upper panel, time-averaged (over $\tau_\alpha/10$, where $\tau_\alpha$ is the 
  structural relaxation time), normalized
  fluctuations of $E$ and $p$ in NVT simulations of the 
  Kob-Andersen\cite{Kob/Andersen:1994} binary 
  Lennard-Jones (KABLJ) system, plotted against time in units of
  $\tau_\alpha \sim 10^3\sigma_{AA}\sqrt{m/\epsilon_{AA}}$. The density was
  1.2$\sigma_{AA}^{-3}$, and the temperature
  $0.474\epsilon_{AA}$. Middle panel, imaginary parts of the three response 
  functions $-c_v(\omega)$, $-\beta_V(\omega)$ and $1/\kappa_T(\omega)$, scaled
  to the maximum value. Lower panel, dynamic  Prigogine-Defay ratio for
  the same simulation. The approach towards unity at frequencies smaller than 
  the loss-peak frequency  ($\sim 1/\tau_\alpha$) is exactly
  equivalent to the correlation between time-averaged quantities shown
  in the upper panel [reproduced from
  Ref.~\onlinecite{Pedersen/others:2008a}].}
\end{figure}

A question that is not actively debated in this research field 
(but see, e.g.,  Refs.~\onlinecite{Schmelzer/Gutzow:2006, Bailey/others:2008a,
Ellegaard/others:2007}), is 
whether a single parameter is enough to describe a highly viscous
liquid. The consensus for more than 30 years is that with few 
exceptions these liquids require more than just one parameter, a conclusion
scarcely surprising given their complexity. 
The meaning of ``having a single parameter'' can be
understood as follows. Following a sudden change in volume, both
pressure and energy relax to 
their equilibrium values over a time scale of minutes or
even hours, sufficiently close to the glass transition. If a single
parameter governs the internal relaxation of the liquid, then
both pressure and energy relax with the same time scale, and in fact
the normalized relaxation functions are identical.\cite{Bailey/others:2008a,
Ellegaard/others:2007} This behavior can
be expressed in the frequency domain, as a certain quantity, the dynamic
Prigogine-Defay ratio, being equal to unity.\cite{Ellegaard/others:2007} 
A key feature of highly viscous liquids is the separation of time scales
between the slow structural (``alpha'') relaxation (up to of order seconds) 
and the very short times (of order picoseconds) characterizing the vibrational 
motion of the molecules.
This separation allows a more direct experimental
consequence of $W,U$ correlations than that described in the previous
subsection: Suppose a highly viscous liquid has perfectly correlated $W,U$
fluctuations. When $W$ and $U$ are time-averaged over, say, one tenth
of the alpha relaxation time $\tau_\alpha$,\cite{Pedersen/others:2008a}
they still correlate
100\%. Since the kinetic contribution to pressure is fast, its time-average
over $\tau_\alpha/10$ is just its thermal average, and thus the
time-averaged pressure equals the time-average of $W/V$ plus a
constant. Similarly, the time-averaged energy equals the
time-averaged potential energy plus a constant. Thus the fluctuations
of the time-averaged $W$ and $U$ equal the slowly fluctuating parts
of pressure and energy, so these slow parts will also correlate 100\% in their
fluctuations. In this way we get from the non-observable quantities
$W$ and $U$ to the observable ones $E$ and $p$ (we similarly
averaged to observe the correlation in 
the SQW system in Paper I). The upper part of Fig.~\ref{combinedPDR} shows
normalized fluctuations of energy and pressure for the commonly studied
Kob-Andersen binary Lennard-Jones system\cite{Kob/Andersen:1994}
(referred to as KABLJ in Paper I), time-averaged over one tenth of 
$\tau_\alpha$. In the lower part we show the
dynamic Prigogine-Defay ratio,\cite{Ellegaard/others:2007} which in the NVT
ensemble is defined as follows:

\begin{equation}
\Lambda_{TV}(\omega) \equiv -\frac{c_V''(\omega)(1/\kappa_T(\omega))''}
{T \left[\beta_V''(\omega)\right]^2};
\end{equation}

\nod here $\kappa_T = 1/K_T$ is the isothermal compressibility and $''$ denotes
 the imaginary part of the complex frequency dependent response function. A way
 to interpret this quantity can be found by considering
the fluctuation-dissipation theorem expressions for the response functions. 
For example the frequency-dependent constant-volume specific heat
 $c_V(\omega)$ is given\cite{Nielsen/Dyre:1996} by

\begin{equation}
c_V(\omega) = \frac{\angleb{(\Delta E)^2}}{k_B T^2} - \frac{i\omega}{k_B T^2}
\int_0^\infty{\angleb{\Delta E(0)\Delta E(t)}\exp(-i\omega t)} dt
\end{equation}

\nod where $E$ is the total energy. Taking the imaginary part we have

\begin{equation}
c_V''(\omega) =  - \frac{\omega}{k_B T^2}(\mathcal{L}\{\angleb{\Delta E(0)\Delta E(t)}\})'
\end{equation}

\nod where we use $\mathcal{L}$ to represent Laplace transformation. Similarly 

\begin{equation}
(1/\kappa_T)''(\omega) = \frac{\omega V}{k_B T}
(\mathcal{L}\{\angleb{\Delta p(0)\Delta p(t)}\})'
\end{equation}

\nod and

\begin{equation}
\beta_V''(\omega) = -\frac{\omega}{k_BT^2}
(\mathcal{L}\{\angleb{\Delta E(0)\Delta p(t)}\})'
\end{equation}

\nod Forming the Prigogine-Defay ratio then gives, after cancelling factors of 
$k_B$, $T$, $V$, and $\omega$,

\begin{equation}
\Lambda_{TV}(\omega) = \frac{(\mathcal{L}\{\angleb{\Delta E(0)\Delta E(t)}\})'
  (\mathcal{L}\{\angleb{\Delta p(0)\Delta p(t)}\})'}
{(\mathcal{L}\{\angleb{\Delta E(0)\Delta p(t)}\})'^2}.
\end{equation}

\nod We can see that the right-hand side has a similar structure to a 
correlation
coefficient, if we take the inverse square root. So in a loose sense the 
dynamic Prigogine-Defay ratio can be thought of as the inverse square of
 a correlation coefficient, referred to a particular time scale. This gives an
intuitive reason for why it is in general greater or equal to
unity, with equality only achieved in the case of 
perfect correlation.\cite{Ellegaard/others:2007} The
lower panel of Fig.~\ref{combinedPDR} shows this quantity for a range
of frequencies for the KABLJ system. It clearly approaches one at low 
frequencies, and stays within 20\% of one in the main relaxation region. In the
sense above, this corresponds to $R>0.9$, or strongly correlation.

 The line of reasoning presented here
 opens for a new way of utilizing computer simulations to understand
 ultraviscous liquids. Present-day computers are barely able to simulate 
one microsecond of real-time dynamics, making it difficult to predict the
 behavior of liquids approaching the laboratory glass transition. We find 
that pressure-energy correlations are almost independent of viscosity, 
however, which makes it possible to make predictions regarding the 
relaxation properties even in the second or hour range of characteristic 
times. Thus if a glass-forming liquid at high temperatures (low viscosity) 
has very strong pressure-energy correlations ($R\sim1$), 
its eight thermoviscoelastic response functions at ultraviscous conditions may 
basically be expressed in terms of 
just one, irrespective of temperature (or viscosity).

\subsection{\label{AgingAndEnergyLandscapes}Aging and energy landscapes}

We now discuss the significance of the present results for the 
interesting results reported in 2002 by Mossa 
{\it et al},\cite{Mossa/others:2002} 
who studied the inherent states visited by the Lewis-Wahnstr\"om
 model\cite{Lewis/Wahnstrom:1994} of the glass-forming liquid
orthoterphenyl (OTP) during aging, i.e., the approach to equilibrium. 
An inherent state (IS) is a local minimum of the so-called
potential energy landscape (PEL) to which a given configuration 
is mapped by steepest-descent minimization.\cite{Goldstein:1969,
Stillinger:1995} The PEL formalism involves modeling the distributions and 
averages of properties of the IS
in the hope of achieving a compact description of the thermodynamics of
glass-forming liquids.\cite{Sciortino:2005, Heuer:2008} The thesis of 
Ref.~\onlinecite{Mossa/others:2002} and of the previous work
\cite{LaNave/Mossa/Sciortino:2002} is that an equation of state can be derived 
using this formalism which is valid even for non-equilibrium situations.
This involves including an extra parameter, namely the average IS (potential) 
energy, $\langle e_{IS} \rangle$, so that the equation of state takes the form

\begin{equation}
p(T, V, \angleb{e_{IS}}) = p_{IS}(\angleb{e_{IS}}, V)
+p_{\textrm{vib}}(T, V, \angleb{e_{IS}})
\end{equation}

\nod where $p_{IS}$ is the ensemble averaged inherent state pressure---for a
given configuration it is the pressure of the corresponding inherent 
state---and $p_{\textrm{vib}}\equiv p-p_{IS}$. The usefulness of splitting in this 
way lies in the fact that $p_{IS}$ does not explicitly depend on $T$.

In equilibrium $\angleb{e_{IS}}$ is determined by $V$ and $T$. The conclusion of
Ref.~\onlinecite{Mossa/others:2002} is that knowledge of $\langle 
e_{IS} \rangle$ in non-equilibrium situations is enough to predict the
corresponding pressure (given also $T$ and $V$). This was based on extensive 
simulations of various aging schedules. Thus the authors conclude that the 
inherent states visited by the system while out of equilibrium, must be in 
some sense the ``same'' ones sampled during equilibrium conditions. ``Same'' is
 effectively 
defined by their results that averages of various inherent state properties
($V$, $e_{IS}$, $p_{IS}$, as well as a
measure of the IS curvature) are all related to each other the same way under
non-equilibrium conditions as under equilibrium conditions. It was
similarly found that the volume could be determined from 
$\langle e_{IS}\rangle$ following a pressure jump in a pressure-controlled
simulation. On the other hand, subsequent work by the same group found that 
this was not at all possible for glassy 
water during compression/decompression cycles.
\cite{Giovambattista/Stanley/Sciortino:2003}

Now, our results (Paper I) for the same OTP model show that it is a strongly
correlating liquid. Thus
 we expect a general correlation between individual, not just average,
 values of $p_{IS}$, the inherent state pressure (which lacks a kinetic term 
and therefore equals the inherent state virial divided by volume) and 
 $e_{IS}$, for a given volume. Therefore, for any given collection of inherent 
states with the same volume---not just equilibrium ensembles---the mean values 
of $U$ and $W$ will fall on the same straight line as the instantaneous values.
Note that this would not hold if the correlation was non-linear.
Correspondingly for given $p_{IS}$, there is a general
correlation between individual values of $e_{IS}$ and $V$. In fact, any two of 
these quantities determine the third with high 
accuracy, and this is true at the level of individual configurations,
including inherent states.

To see how this works for cases involving fixed volume, we write the total 
(instantaneous) pressure as a sum of an inherent state part, which involves the
 the virial at the corresponding inherent state, plus a term involving the 
difference of the true virial from the inherent state virial, plus the kinetic 
term:

\begin{equation}
p = \frac{W_{IS}}{V} + \frac{W-W_{IS}}{V} + \frac{N k_B T}{V}.
\end{equation}

\nod The first term is linearly related to the inherent state energy for a
strongly correlating liquid. Moreover, the difference term is similarly
expressed in terms of the corresponding energy difference,
 $W-W_{IS}=\gamma(U-e_{IS})$. Taking averages over the (possibly 
non-equilibrium, although we assume equilibrium within a given
potential energy basin) ensemble, we expect that
$\langle U-e_{IS}\rangle$ depends only
on $T$ and $V$ (a slight $e_{IS}$
dependence can appear in $\gamma$ since this is slightly 
state-point dependent). Thus it follows that $p$ can be reconstructed from a
knowledge of (average) $e_{IS}$, $V$ and $T$, without any assumptions about the
nature of the inherent states visited. In particular, no conclusion can be 
drawn regarding the latter. The failure of the pressure 
reconstruction in the case of 
water~\cite{Giovambattista/Stanley/Sciortino:2003} is not
surprising, since water models are generally not 
strongly correlating (which as we saw in Paper I is 
linked to the existence of the density maximum).

\subsection{\label{Biomembranes}Biomembranes}

A completely different area of relevance for the type of correlations
reported here relates to recent work of Heimburg and 
Jackson,\cite{Heinburg/Jackson:2005} who have proposed a
controversial new theory of nerve
signal propagation. Based on experiment and theory they suggest that
a nerve signal is not primarily electrical, but a soliton 
sound-wave.\cite{Heinburg/Jackson:2007b} Among
other things this theory explains how anaesthesia works (and why one
can dope people with the inert gas Xenon): Anaesthesia works simply by
a freezing-point depression that changes the membrane phase transition
temperature and affects its ability to carry the soliton sound wave. A
crucial ingredient of the theory is the postulate of proportionality
between volume and enthalpy of microstates, i.e., that their thermal
equilibrium fluctuations should correlate perfectly. This should apply
even through the first-order membrane {\em melting}
transition. The theory was justified in part from previous experiments by
Heimburg showing proportionality between compressibility and specific
heat through the phase transition.\cite{Ebel/Grabitz/Heimburg:2001} 
The postulated correlation---including the claim that is survives a 
first-order phase transition---fits precisely
the pattern found in our liquid simulations. 


\begin{table}
\caption{Data from NpT simulations of fully hydrated phospholipid membranes of 
1,2-Dimyristoyl-sn-Glycero-3-Phosphocholine (DMPC),  
1,2-Dimyristoyl-sn-Glycero-3-Phospho-L-Serine with sodium as counter ion 
(DMPS-Na), hydrated DMPS (DMPSH), and 
1,2-Dipalmitoyl-sn-Glycero-3-Phosphocholine (DPPC).\cite{Pedersen/others:2008b,
Pedersen/others:2008c} The columns list 
temperature, correlation coefficient between volume and energy, average lateral
area per lipid, simulation time in equilibrium, and total simulation time.}
\label{membraneSimulations}
\begin{tabular}{lccccc}
\hline
 & $T$ [K] & $R_{EV}$ & $A_{lip}$ [\AA{}$^2$] & $t$ [ns] & $t_{tot}$ [ns]\\
 \\
\hline
DMPC     & 310  & 0.885  & 53.1 & 60 & 114  \\
DMPC     & 330  & 0.806  & 59.0 & 50 & 87   \\
DMPS-Na  & 340  & 0.835  & 45.0 & 22 & 80   \\
DPPC     & 325  & 0.866  & 67.3 & 13 & 180  \\
\hline
\end{tabular}
\end{table}

By re-examining existing simulation 
data\cite{Pedersen/others:2006b,Pedersen/Peters/Westh:2007} as well as carrying
out extensive new simulations,\cite{Pedersen/others:2008b} we have investigated
 whether the correlations are also found in several model membrane systems, 
five of which are listed in Table~\ref{membraneSimulations}. The
simulations involved a layer of phospholipid membrane surrounded by water, in
the $L_\alpha$ phase
(that is, at temperatures above the transition to the gel-state), at
constant $p$ and $T$. When $p$, rather than $V$, is constant, the relevant
 quantities that may correlate are energy and volume. As with viscous liquids 
(subsection~\ref{ViscousLiquids}) and the square well system 
(Paper I), time-averaging is necessary for a 
correlation to emerge, where now

\begin{equation}
\Delta \bar{E(t)} \simeq \gamma^{vol} \Delta \bar{V(t)}
\end{equation}

\begin{figure}
\includegraphics[width=3.5in]{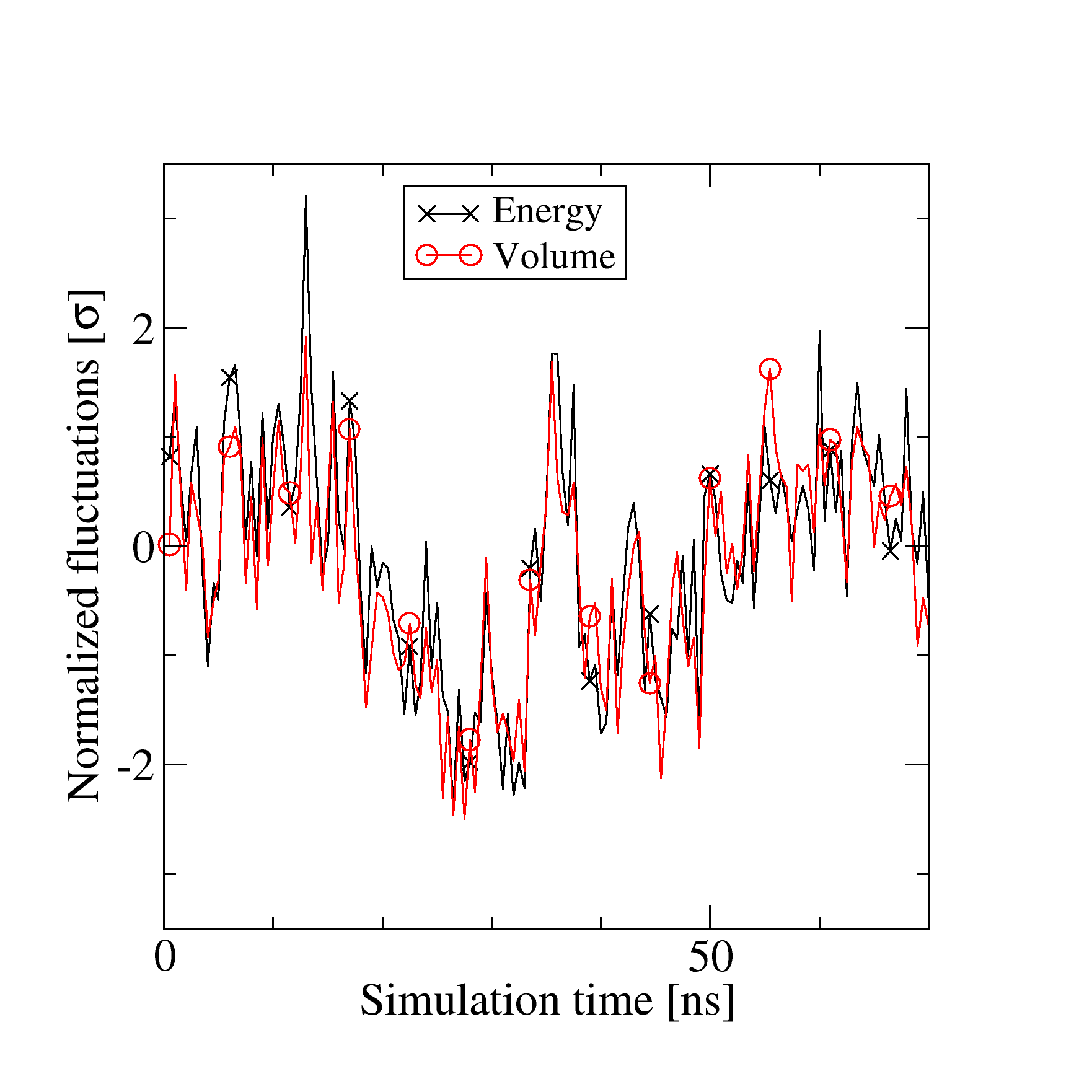}
\caption{\label{membraneTimeSeries} (color online) Normalized fluctuations of
energy ($\times$) and volume ($\circ$) for a 
1,2-Dipalmitoyl-sn-Glycero-3-Phosphocholine  (DPPC) membrane at 325 
K.\cite{Pedersen/others:2008c} Each data point represents a 0.5 ns average. 
Energy and volume are correlated with a correlation coefficient of $R=0.87$
(NpT).}
\end{figure}

When an averaging time of 1 ns is chosen, a significant correlation emerges, 
with correlations between 0.8 and 0.9 (Table~\ref{membraneSimulations}; note
these $R_{EV}$ values fall slightly short of our criterion of 0.9 for
 ``strong correlations''). The time series of time-averaged, normalized $E$ and
 $V$ for one case are shown in Fig.~\ref{membraneTimeSeries}. The necessity of
 time-averaging stems from the presence of water, which we know does not 
exhibit strong correlations. Since the membrane dynamics are much slower than 
those of the water, they can be isolated by time averaging.

\section{\label{Conclusion}Conclusions and outlook}

In Paper I and this work we have demonstrated several cases of strongly 
correlating liquids and some cases where the correlation is weak or absent 
(at least under normal conditions of pressure and temperature). An important
next step is to continue to document the existence or
 non-existence of correlations, particularly in different kinds of 
model systems, such as non-pair potential systems and systems with interactions
computed using quantum mechanics (e.g., by density functional theory). It is 
noteworthy in this respect that after suitable time averaging 
the correlations may appear in systems where 
they are otherwise unexpected. One example was the square-well (SQW) case, 
where the correlation was between the time-averaged virial and potential 
energy. In the case of viscous liquids time averaging allowed a correlation to
appear between the more accessible energy and pressure, while for the
biomembranes it made it possible to remove the non strongly-correlating
contributions from the water. In all cases time averaging is relevant because 
of a separation of time scales: the square-well system because the time scale 
over which the average number of neighbors changes is long compared to the time
 between collisions; viscous liquids because the vibrational dynamics (which
includes the kinetic contributions) is fast compared to the slow 
configurational dynamics; biomembranes because the membrane dynamics is slow 
compared to those of the water. Note that in this case it was necessary
to consider an NpT ensemble and study correlations between energy and volume,
because only a part of the system is strongly correlating, and this part cannot
be constrained to a particular volume. It is worth noting that even with fixed
volume, the correlation coefficient depends on whether the ensemble is NVT or
 NVE, although the strongly correlating limit of $R\rightarrow 1$ is 
independent of ensemble.

A point which has been mentioned, but which is
worth emphasizing again, is that the replacement of the potential by an
appropriate inverse power law can only reproduce the fluctuations, and not the
mean-values of (potential) energy and virial, nor their first derivatives with
respect to $T$ and $V$. These determine the equation of state, in particular
 features such as the van der Waals loop which are absent in a pure power-law
system, even if changes in the exponent are allowed. 
The generalization to the extended effective inverse power-law 
approximation, however, allows in principle such features to be described. We 
are currently investigating the dependence of the parameters of the extended
approximation on state point.

Finally, consider again viscous liquids, which are typically deeply 
supercooled. The most common way of classifying them involves the fragility
 parameter introduced by Angell,\cite{Angell:1985}
 related to the departure from Arrhenius 
behavior of the temperature dependence of the viscosity. Strong liquids,
having the most Arrhenius behavior, have traditionally been considered the
easiest ones to understand, because Arrhenius temperature dependence is
 well-understood. But it may well be that
 strongly correlating liquids, are in fact the 
simplest.\cite{LeGrand/others:2007}
The connection with the long-discussed question of whether a single-order 
parameter describes highly viscous liquids has been discussed briefly in 
subsection~\ref{ViscousLiquids}, and is
discussed further in Ref.~\onlinecite{Bailey/others:2008a}. As an example,
a direct application of the ``strongly-correlating'' property concerns
diffusion in supercooled liquids. Recent work by Coslovich and Roland 
\cite{Coslovich/Roland:2008} has shown that the diffusion constant
in viscous binary Lennard-Jones mixtures may be fitted by an expression 
$D=F(\rho^\gamma/T)$ where $\gamma$ reflects the effective inverse power of 
the repulsive core. ``Density scaling'' has also been observed 
experimentally.\cite{Tarjus/others:2004,Casalini/Roland:2004,
Dreyfus/others:2004,Roland/others:2005} It is natural, given Coslovich and 
Roland's results,
to hypothesize that the scaling exponent is connected to pressure-energy
correlations, and in Ref.~\onlinecite{Pedersen/others:2008}
it was conjectured that density scaling applies if and only if
the liquid is strongly correlating. We have very recently studied the 
relationship between the two quantitatively\cite{Schroder/Pedersen/Dyre:2008a} 
and have found that (1) 
density scaling does indeed hold to the extent that the liquids are strongly 
correlating, and (2) the scaling exponent is given accurately by the slope 
$\gamma$ of the correlations (hence our use of
the same symbol). This finding supports the conjecture that
strongly correlating liquids may be simpler than liquids in general.

In summary, the property of strong correlation between the equilibrium 
fluctuations of virial and potential energy allows a new way to classify
 liquids. It is too soon to tell how fruitful this will turn out in the long 
term, but judging from the applications briefly presented here, it seems at 
least plausible that it will be quite useful.

\begin{acknowledgments}
Useful discussions with S{\o}ren Toxv{\ae}rd are gratefully acknowledged.
Center for viscous liquid dynamics ``Glass and Time'' is sponsored by The 
Danish National Research Foundation.
\end{acknowledgments}
















































\end{document}